\documentclass[draftclsnofoot,onecolumn,11pt,conference]{IEEEtran}
\usepackage{setspace}

\usepackage{cite}
\usepackage{graphicx}
\usepackage{psfrag}
\usepackage{subfigure}
\usepackage{amsmath}
\usepackage{amssymb}
\usepackage{epsfig}
\usepackage{epsf}

\newtheorem{theorem}{Theorem}

\newtheorem{corollary}[theorem]{Corollary}

\newtheorem{lemma}[theorem]{Lemma}

\def\bb0{{\mathbb{0}}}

\def\bb{{\mathbf{b}}}

\def\bff{{\mathbf{f}}}
\def\bg{{\mathbf{g}}}
\def\bh{{\mathbf{h}}}

\def\bw{{\mathbf{w}}}
\def\bx{{\mathbf{x}}}
\def\by{{\mathbf{y}}}

\def\b0{{\mathbf{0}}}

\def\bI{{\mathbf{I}}}

\def\bR{{\mathbf{R}}}

\def\bW{{\mathbf{W}}}
\def\bX{{\mathbf{X}}}

\def\bbC{{\mathbb{C}}}

\def\bbE{{\mathbb{E}}}

\def\cH{\mathcal{H}}

\def\cN{\mathcal{N}}

\def\cW{\mathcal{W}}

\def\sfee{{\mathsf{e}}}

\def\sf0{{\mathsf{0}}}

\def\Nt{{N_t}}

\def\Bt{{B_{\rm tot}}}
\def\Bd{{B_{k,{\rm (d)}}}}
\def\Bi{{B_{k,{\rm (i)}}}}
\def\Td{{T_{k,{\rm (d)}}}}
\def\Ti{{T_{k,{\rm (i)}}}}

\def\ln{{\rm ln}}

\begin{document}

\title{\LARGE Adaptive Limited Feedback for Sum-Rate Maximizing Beamforming in Cooperative Multicell Systems}
\author{\authorblockN{Ramya Bhagavatula and Robert W. Heath, Jr.}
\authorblockA{Department of Electrical \& Computer Engineering \\
Wireless Networking and Communications Group\\
The University of Texas at Austin\\
1 University Station C0806 \\
Austin, TX 78712-0240 \\
\{bhagavat, rheath\}@ece.utexas.edu\\}}

\maketitle

\begin{abstract}
Base station cooperation improves the sum-rates that can be achieved in cellular systems. Conventional cooperation techniques require sharing large amounts of information over finite-capacity backhaul links and assume that base stations have full channel state information (CSI) of all the active users in the system. In this paper, a new limited feedback strategy is proposed for multicell beamforming where cooperation is restricted to sharing only the CSI of active users among base stations. The system setup considered is a linear array of cells based on the Wyner model. Each cell contains single-antenna users and multi-antenna base stations. Closed-form expressions for the beamforming vectors that approximately maximize the sum-rates in a multicell system are first presented, assuming full CSI at the transmitter. For the more practical case of a finite-bandwidth feedback link, CSI of the desired and interfering channels is quantized at the receiver before being fed back to the base station. An upper bound on the mean loss in sum rate due to random vector quantization is derived. A new feedback-bit allocation strategy, to partition the available bits between the desired and interfering channels, is developed to approximately minimize the mean loss in sum-rate due to quantization. The proposed  feedback-bit partitioning algorithm is shown, using simulations, to yield sum-rates close to the those obtained using full CSI at base stations. 
\end{abstract}

\newpage

\section{Introduction} \label{sec:Intro}
Base station cooperation is an effective strategy to increase data rates and reduce outages in multiple-input multiple-output (MIMO) cellular systems \cite{Simeone2009, Ali2009, Choi2008, Shamai2001a, Zhang2004,Jafar2004, Kang2006,Jing2008, Ng2005, Ng2008, Somekh2006,Ekbal2005, Lee2008, Lee2009, Chae2009,Somekh2007}. Cooperative encoding at the base stations can be used to combat co-channel interference (CCI), paving the way for more aggressive frequency reuse, which can lead to higher data rates and simpler network configurations. Consequently, cooperative transmission is being considered for upcoming cellular standards like 3GPP long term evolution advanced that are targeting universal frequency reuse \cite{3GPPAdv}. Base station cooperation entails sharing control signals, user propagation channel information and/or precoding data via high-capacity wired backhaul links to coordinate transmissions \cite{Simeone2009, Ali2009, Choi2008, Shamai2001a, Zhang2004, Jafar2004, Kang2006,Jing2008, Ng2005, Ng2008, Somekh2006, Ekbal2005, Lee2008, Lee2009, Chae2009, Somekh2007}. Most of the literature on cooperative techniques assumes that base stations have full channel state information (CSI) of all active users in the system. This is infeasible as receivers use finite-bandwidth channels to feedback CSI to base stations (in frequency division duplex systems) \cite{Love2008}. 
Another common assumption made in multicell cooperative literature is that base stations are connected using ideal backhaul links. In practical systems, however, the prohibitive cost of connecting all the base stations in the network using high-capacity links restricts backhaul capabilities \cite{Simeone2009}. While sharing a larger amount of information among base stations can improve performance, the extent of cooperation among base stations is restricted by the capacity-limited backhaul. Hence, it is important to develop cooperative techniques with limited feedback that maximize performance gains while ensuring a manageable load on the finite-capacity backhaul. 

Perhaps the least aggressive form of cooperation is joint resource allocation or joint scheduling by base stations in adjacent cells \cite{Ali2009, Choi2008}. These strategies typically result in a small load on the backhaul link and possess comparatively low complexity. For example, in dynamic and fractional frequency reuse, base stations exchange control-level information to cooperatively assign different frequency bands of operation to users in adjacent cells \cite{Ali2009}. By utilizing the available spectrum more efficiently, the techniques in \cite{Ali2009} yield higher sum-rates as compared to static frequency reuse, which does not account for varying user traffic. Alternatively, joint inter-cell scheduling assigns different transmission cycles to users in adjacent cells using opportunistic scheduling \cite{Choi2008}, introducing an expanded multiuser diversity gain relative to static frequency reuse. Frequency reuse and inter-cell scheduling do not utilize all the available frequency and time resources, respectively and hence, do not realize the performance gains that can be potentially obtained using base station cooperation \cite{Choi2008}.

Full cooperation among base stations leads to the highest sum-rates, at the cost of increased overhead associated with the exchange of a greater amount of information among base stations. The sum-capacity maximizing solution was derived and found to be achieved by the dirty paper coding (DPC) \cite{Costa1983, Shamai2001a, Zhang2004, Jafar2004}. DPC requires joint precoder design, where all the base stations have perfect knowledge of the interference seen by each user. DPC was proposed for single antenna cellular systems in \cite{Shamai2001a}. Multicell DPC using MIMO was investigated in \cite{Zhang2004}, and an upper bound was derived on the sum-rates achievable using base station cooperation. While the approaches in \cite{Shamai2001a, Zhang2004} assumed a sum-power constraint across all the base stations, DPC with more realistic per-base power constraints was considered in \cite{Jafar2004}. Multicell DPC is difficult to implement in practice due to the requirement for the base stations to have a non-causal knowledge of the interference \cite{Zhang2004, Kang2006}. As a result, several low-complexity and relatively more practical joint transmission strategies were proposed in \cite{Zhang2004,Jing2008} using full cooperation and per-base station power constraints. These sub-optimal linear techniques, including zero forcing, minimum mean square error or null-space decomposition, yield sum-rates that are far from that of multicell DPC due to the transmit power inefficiency introduced by the precoding matrices, especially for low-rank channels \cite{Zhang2004,Jing2008}. Though the sub-optimal methods reduce the encoding complexity, they still require full CSI at the base stations and involve the exchange of a large amount of information among base stations, resulting in a prohibitive load on the finite-capacity backhaul links \cite{Simeone2009,Zhang2004, Jing2008}. 

Partial cooperative strategies, where base stations exchange only the CSI of active users, offer a fair balance between ensuring a reasonable load on the backhaul links and attaining the performance gains using cooperation \cite{Simeone2009}. The shared CSI can be used by the base stations to design individual precoding matrices (or beamforming vectors, for single-stream transmission) on site to transmit exclusively to users within their own cell \cite{Shamai2001a, Zhang2004, Jafar2004}. Most of the existing literature \cite{Ng2005, Ng2008,Somekh2006,Ekbal2005,Lee2008,Lee2009,Chae2009} also assumes that users can feedback full CSI to the base stations. A distributed approach for beamforming was studied in \cite{Ng2005, Ng2008} where transmit symbol vectors were designed using linear minimum mean square error (MMSE) estimation techniques. The forward-backward and the sum-product Kalman smoothing algorithms were used to iteratively obtain the transmit vector as an optimal linear MMSE estimate for a linear and a 2-D array of cells, respectively \cite{Ng2005,Ng2008}. The convergence of these iterative approaches is, however, not guaranteed. A distributed zero-forcing beamforming strategy was proposed in \cite{Somekh2006} using a scheduling algorithm to select users with the best desired channels in each cell. The sub-optimal approach is not feasible for practical setups as it uses a sum-power constraint across all base stations, and only satisfies an equal power per base station asymptotically for a large number of users per cell \cite{Somekh2006}. Distributed transmit beamforming was also investigated in \cite{Ekbal2005}, where the authors propose an iterative algorithm to minimize the transmit power that does not necessarily maximizie the sum-rates. The authors of \cite{Lee2008,Lee2009,Chae2009} propose a beamforming solution to approximately maximize the sum-rates for only a two-cell system.


In this paper, we propose a linear beamforming strategy to approximately maximize the sum-rates at high signal-to-interference noise ratios (SINR) in multicell systems, using partial cooperation. The proposed beamforming method is similar to the two-cell approaches in \cite{Lee2008,Lee2009,Chae2009}. The cellular setup considered is based on the Wyner model \cite{Wyner1994}, where neighboring base stations share only the CSI to beamform independently to one active user in the cell with a single (dominant) co-channel interferer. The proposed algorithm is also applicable to a finite linear array of cells. The variation in the strength of the desired and interfering signals is taken into account to maximize the sum-rate using a generalized eigenvector approach. We first assume full CSI availability at base stations and a high-capacity backhaul to arrive at the closed-form expressions for the \emph{mutlicell} beamforming vectors. We then consider the more realistic limited feedback and capacity-limited backhaul in the rest of the paper. The beamforming strategy presented is non-iterative in nature and hence, does not have convergence issues unlike the solutions in \cite{Ng2005, Ng2008}. The linear approach in this paper uses explicit per-base station power constraints, in contrast to distributed zero-forcing beamforming in \cite{Somekh2006}. Further, it possesses low-complexity, and results in a smaller burden on the backhaul link, in comparison to the full cooperation strategies in \cite{Shamai2001a, Zhang2004, Jafar2004, Jing2008}, while yielding sum-rates that are reasonably close to those of multicell DPC. Also, it can be applied to models with any number of linearly arranged cells, and not just two-cell situations as in \cite{Lee2008,Lee2009,Chae2009}.

Implementing the linear beamforming strategy requires full CSI at the base stations to design the transmit beamforming vectors, which may be impractical. Further, all the strategies described in \cite{Shamai2001a, Zhang2004,Jafar2004, Kang2006,Jing2008, Ng2005, Ng2008, Somekh2006, Ekbal2005, Lee2008, Lee2009, Chae2009,Simeone2009, Somekh2007} require full CSI at the base stations. In practice, as feedback links have finite bandwidth, quantized CSI is fed back to the base stations using the concept of limited feedback \cite{Love2008}. While limited feedback for single-cell systems is well researched (refer to \cite{Love2008,Mielczarek2008} and the references within), comparatively less work has been done in the multicell scenario. In contrast to the single-cell case, the CSI of multiple channels needs to be fed back for the cooperative-based strategies. Hence, it is not even clear how to apply the existing multiuser limited CSI feedback techniques \cite{Zhang2009a, Trivellato2008} to multicell systems. Also, the varying strengths of the desired and interfering channels need to be taken into account when developing a multicell limited feedback strategy that will efficiently utilize feedback resources. Single-cell multiuser limited feedback does not have to deal with the variability in signal strengths, due to the single-channel feedback. Note that the existing work in multicell cooperation that considers the difference in signal strengths of the desired and interfering channels \cite{Jing2008,Somekh2006,Somekh2007} assumes full CSI at base stations. 

In this paper, we develop a limited feedback strategy for the proposed beamforming algorithm that will partition bits between the desired and interfering channels as a function of their relative strengths. For analytical reasons, the desired and interfering channels are quantized using random vector quantization (RVQ), i.e. the quantization vectors are independently chosen from the isotropic distribution on a unit hypersphere \cite{Santipach2004, Santipach2005}. We first present a simple model for limited feedback of CSI in multicell systems. The model requires each user to feedback quantized CSI to its own base station. Adjacent base stations are connected by a backhaul link, which is used to transfer the quantized CSI of only the interfering channels. Hence, the requirement of a global high-capacity backhaul link is eliminated. We quantify the reduction in sum-rate caused by quantizing CSI using RVQ by deriving an upper-bound on the mean loss in sum-rate. We also present a feedback-bit allocation technique that approximately minimizes the mean loss in sum-rate and hence, utilizes the available feedback resources efficiently. Using the proposed algorithm, we show that the sum-rates approach those of multicell DPC even for a limited feedback scenario.

We first extend the two-cell cooperative beamforming approach in \cite{Lee2008,Lee2009, Chae2009} to a multicell system where adjacent base stations exchange only CSI to approximately maximize sum-rate. The main contributions of this paper are as follows.
\begin{itemize}
\item We propose a novel and simple limited feedback model for the multicell system, where each user feeds back quantized CSI of the desired and interfering channels to its own base station. In the proposed model, the neighboring base stations exchange only the quantized interfering CSI. This makes the load on the backhaul link manageable. By ensuring that only the adjacent base stations are connected, the requirement of a globally connected backhaul is eliminated.
\item We analyze the performance of the beamforming technique presented in this paper with limited feedback by deriving an upper bound for the mean loss in sum-rate due to CSI quantization using RVQ. It is shown via simulations, that the upper bound is reasonably tight.
\item We present a new feedback-bit partitioning algorithm to allocate the available feedback bits between the desired and interfering channels to approximately minimize the mean loss in sum-rate due to limited feedback.
\end{itemize}

This paper is organized as follows. We first describe the system models used in this paper in Section \ref{sec:SysModel}. We then present the multicell beamforming strategy, and compare it with two existing beamforming techniques for the full CSI case in Section~\ref{sec:BFVec_FullCSI}. In Section~\ref{sec:BFVec_PartialCSI}, we consider the limited feedback scenario for the beamforming algorithm in Section~\ref{sec:BFVec_FullCSI}. We propose a feedback-bit allocation strategy to partition the available bits between the desired and interference channels in Section~\ref{sec:OptFB_PartialCSI}. The simulation results verifying the accuracy of the limited feedback algorithm are presented in Section~\ref{sec:Results}. The concluding remarks are provided in Section~\ref{sec:Conclusion}, and detailed proofs are presented in Appendices~\ref{app:meanLogCos} - \ref{app:Delta_Convexity} .   

\noindent\textbf{Notation:} In this paper, $\bX$ refers to a matrix and $\by$ stands for a vector. The transpose and conjugate of $\bX$ are given by $\bX^T$ and $\bX^c$, respectively. The Hermitian transpose of $\bX$ is given by $\bX^*$. The inverse and pseudo-inverse of $\bX$ are given by $\bX^{-1}$ and $\bX^\dag$, respectively. An identity matrix of size $R \times R$ is denoted by $\bI_R$. $\bbE\{.\}$ refers to the expectation. $\|\bx\|$ stands for the Frobenius norm of $\bx$. $\cN_c(\mu,\sigma)$ refers to a complex Gaussian random distribution with mean $\mu$ and variance $\sigma$. 

\section{System Model} \label{sec:SysModel}
Consider the multicell setup shown in Fig.~\ref{fig:WynerModel} for an array of linearly arranged cells. The multicell system used in this paper is based on the Wyner model \cite{Wyner1994}. We assume that the base station in each cell serves a single active user \cite{Shamai2001a}, using intra-cell time division multiple access. Each user is assumed to face interference from \emph{one} of its neighboring cells, as shown in Fig.~\ref{fig:WynerModel}. The received signal power of the desired and interfering signals is a function of the user's location in the cell. A similar approach was adopted in \cite{Jing2008}, where a user at the cell center receives a signal from only its corresponding base station\footnote{In our paper, we assume that the interference from the adjacent cell is negligible at the cell center.}, while a user at the cell edge is subjected to interference from an adjacent cell as well. The results in this paper can also be applied to modifications of the Wyner model, including the circular arrangement of cells proposed in \cite{Somekh2006}, and a finite linear cellular array. Note that the circular extension is a generalization of the two-cell case and is equivalent to the Wyner model as the number of cells becomes large. The generalized eigenvector beamforming strategy proposed in this paper can be extended to the finite-array case by accounting for the edge effects, as will be described in Section~\ref{sec:BFVec_FullCSI}. The Wyner model and its extensions are widely used in literature as they model the central factors of a cellular system like fading and inter-cell interference, while retaining analytical tractability \cite{Jing2008,Somekh2006, Somekh2007}. We assume that the backhaul link used for information exchange between base stations is ideal and that the time delay associated with feedback and cooperation is zero.


The number of cells in the multicell system is denoted by $K$, where $K$ goes to infinity for the Wyner model. We index the users in each cell by the base station they obtain their desired signal from, i.e. the $k^{\rm th}$ base station services the $k^{\rm th}$ user, for $k=1,\ldots,K$. We assume that all the base stations are equipped with $\Nt$ antennas, while each user supports a single receive antenna (i.e. multiple-input single-output, or MISO, system). The channel corresponding to the desired signal between the $k^{\rm th}$ base station and $k^{\rm th}$ user is denoted by $\bh_k\in \bbC^{\Nt \times 1}$. The interfering channel between the $k^{\rm th}$ user and the $(k+1)^{\rm th}$ base station is given by $\bg_{k+1} \in \bbC^{\Nt \times 1}$. This is illustrated in Fig.~\ref{fig:WynerModel}. The symbol transmitted from the $k^{\rm th}$ base station (intended for the $k^{\rm th}$ user) is denoted by $s_k$, where the transmit power, $\bbE\{|s_k|^2\}$ is normalized to one. The transmitted signals are subjected to large-scale fading, which includes distance-dependent path-loss and shadowing effects, and small-scale fading. After averaging over the small-scale fading effects, the desired and interfering signal powers received at the user terminal are denoted by $\gamma_{k,{\rm (d)}}$ and $\gamma_{k,{\rm (i)}}$, respectively, for the $k^{\rm th}$ user. To facilitate analysis in different loss settings, we let $\gamma_{k,{\rm (i)}} = \alpha_k \gamma_{k,{\rm (d)}}$, where $\alpha_k \in [0,~1]$ (i.e. the interfering signal strength can at most be equal to that of the desired signal). Note that a similar parameter is used in \cite{Jing2008} to model the SNR of the interfering signal with respect to the received signal. Using the narrowband flat-fading model, the baseband discrete-time input-output relation for the user in the $k^{\rm th}$ cell is given by\footnote{We drop the discrete-time index for sake of convenience.}
\begin{equation}
y_k = \sqrt{\gamma_{k,{\rm (d)}}} \bh_{k}^T \bff_{k} s_{k} + \sqrt{\gamma_{k,{\rm (i)}}}\bg_{k+1}^T\bff_{k+1} s_{k+1} + n_k ,\label{eqn:InpOutReln}
\end{equation}
where $y_k \in \bbC$ is the received signal at the $k^{\rm th}$ user and $\bff_k\in \bbC^{\Nt \times 1}$ is the beamforming vector at the $k^{\rm th}$ base station. Finally, $n_k \in \bbC$ is complex additive zero-mean white Gaussian noise at the receive antennas, with $\bbE\{|n_k|^2\} = N_o$.


The signal to interference noise ratio (SINR) of the $k^{\rm th}$ user is given by
\begin{eqnarray}
{\mathsf {SINR}}_k &=& \frac{\gamma_{k,{\rm (d)}}|\bh^T_{k}\bff_{k}|^2}{\alpha_k \gamma_{k,{\rm (d)}}|\bg^T_{k+1}\bff_{k+1}|^2 + N_o}
\label{eqn:SINR1}\\
&=& \frac{|\bh^T_{k}\bff_{k}|^2}{\alpha_k|\bg^T_{k+1}\bff_{k+1}|^2 + \frac{1}{\rho_{k,{\rm (d)}}}} \ .\label{eqn:SINR2}
\end{eqnarray}
Note that $ \rho_{k,{\rm (d)}} = \frac{\gamma_{k,{\rm (d)}}}{N_o}$ is the received SNR of the desired signal, which is independent of the beamforming vectors. The base stations are assumed to have perfect knowledge of $\rho_{k,{\rm (d)}}$. This is a popular assumption in literature \cite{LF-Assumption1,LF-Assumption2,Huang2009}. It was shown in \cite{Huang2009} that SNR quantization does not effect the sum-rates of a single-cell multiuser MIMO system signiÞcantly. To the best of our knowledge, the effect of SNR quantization on the sum-rates of a multicell system has not yet been investigated. We analyze the impact of SNR quantization in our future work on cooperative beamforming. In this paper, however, we assume that the base stations have perfect knowledge of the SNR so we can concentrate on quantizing the channel direction. The sum-rate of all the users within the system, $R_s$, is expressed as 
\begin{equation}
R_s = \sum_k \log_2\left( 1 + {\mathsf{SINR}}_k \right) \ .
\label{eqn:SumRate}
\end{equation}
It is evident that the design of the beamforming vectors  $\left\{\bff_k\right\}_{k=1}^{K}$ influences the SINRs obtained at each of the users and hence, the sum-rate of the multicell system. The dependence of $\mathsf{SINR}_k$ on both, $\bff_k$ and $\bff_{k+1}$ (as given in \eqref{eqn:SINR2}) implies that a joint optimization across all the active users in the multicell system is required to maximize the sum-rate. In this paper, we use a high SINR approximation to remove this interdependency of users and thereby avoid an explicit joint maximization.

\section{Designing Beamforming Vectors Assuming Full CSI} \label{sec:BFVec_FullCSI}
In this section we present a multicell cooperative beamforming approach. First, we describe two popular existing beamforming strategies, non-cooperative eigen-beamforming and cooperative zero-forcing beamforming, which we use for comparison with the proposed algorithm in Section \ref{sec:Results}. Eigen-beamforming maximizes the desired signal strength, while zero-forcing beamforming nulls out the interference. Next, we describe the proposed generalized eigenvector based beamforming strategy that approximately maximizes the sum-rate at high SINR, in a multicell system. We consider the multicell setup given in Section \ref{sec:SysModel} and assume that full CSI is available at the base stations.

\subsection{Eigen-Beamforming and Zero-Forcing Beamforming }\label{sec:ExistingStrategies-FullCSI}
In eigen-beamforming, each base station transmits in a non-cooperative manner by ignoring the co-channel interference from neighboring base stations, as in the case of single-cell single-user beamforming \cite{Mondal2004}. For MIMO systems, the beamforming vector is chosen to be the eigenvector corresponding to the maximum eigenvalue of the MIMO channel. For a MISO system, the eigen-beamforming vector, $\bff_k$, is reduced to 
\begin{equation}
\bff_k = \frac{\bh_k^c}{\|\bh_k\|} \label{eqn:MaxSNR-BFF} \ .
\end{equation}
Note that \eqref{eqn:MaxSNR-BFF} maximizes the numerator in \eqref{eqn:SINR2}. Eigen-beamforming maximizes the data rate that can be obtained in single-cell single-user systems, with no inter-user or inter-cell interference.  

The zero-forcing approach in \cite{Zhang2004} requires base stations to share data in addition to CSI. To ensure fair comparison with the beamforming strategy presented, we implement zero-forcing with single-cell processing in a way that only CSI is shared between base stations, using a principle similar to that in \cite{Zhang2004}. Adjacent base stations exchange the interfering channel state information between each other, so that the $k^{\rm th}$ base station has the knowledge of both $\bh_k$ and $\bg_k$. Assuming full CSI at the transmitter, the multicell zero-forcing criterion requires that $|\bg_k^T\bff_k|^2 = 0$, where $\bff_k^*\bff_k = 1$. Denoting $\bW_k = [\bh_k^T ,~\bg_k^T]^\dag$, the zero forcing beamforming vector is designed by setting
\begin{eqnarray}
\bff_k & = & \frac{\bW_k(:,1)}{\|\bW_k(:,1)\|} \label{eqn:BF-IntNull1}.
\end{eqnarray}
Note that $\bg_k^T\bW_k(:,1) = 0$, and hence, the solution in \eqref{eqn:BF-IntNull1} minimizes the denominator in \eqref{eqn:SINR2}.

\subsection{Proposed Beamforming Strategy for Approximately Maximizing Sum-Rate at High SINR} \label{sec:SINRMax}
In this section, we present a linear multicell cooperative beamforming strategy (for the setup in Section~\ref{sec:SysModel}), based on the \emph{two-cell} beamforming solution proposed in  \cite{Lee2008, Lee2009, Chae2009}. The approach in \cite{Lee2008, Lee2009} approximately maximizes the sum-rate for a two-cell MISO system where base stations are assumed to have full CSI. In this paper, we extend the analysis to develop a cooperative beamforming technique that approximately maximizes the sum-rates in the high SINR regime in a multicell system.

At high SINR, since $\log(1 + \mathsf{SINR})\approx \log(\mathsf{SINR})$, \eqref{eqn:SumRate} can be expressed as
\begin{equation}
R_s \approx \sum_k \log_2\left({\mathsf{SINR}}_k  \right) = \log_2\left(\prod_k {\mathsf{SINR}}_k \right) \label{eqn:SumRate2} \ .
\end{equation}
Thus, maximizing the sum-rate, $R_s$, approximately at high SINR involves maximizing the product of the SINRs\footnote{We show, using simulations, in Section~\ref{sec:Results} that the approximation is tight for small $\alpha_k$ or large $\rho_{k,{\rm (d)}}$.}. The beamforming vectors $\left\{\bff_k\right\}_{k=1}^{K}$ are found by solving
\begin{align}
 R_{s,{\rm opt}} = \max_{\left\{\bff_k\right\}_{k=1}^{K}} ~~ \log_2\left(\prod_{k=1}^K \frac{|\bh_{k}^T\bff_{k}|^2}{\alpha_k|\bg_{k+1}^T\bff_{k+1}|^2 + \frac{1}{\rho_{k,{\rm (d)}}}}  \right)~~ \rm{s.t.}~~\|\bff_k\|^2 = 1,~{\rm for}~k = 1,\ldots,K \label{eqn:C-Optimization1} \ .
\end{align}
Taking advantage of the commutativity of the multiplication operation, the objective function 
can be written as
\begin{align}
R_{s,{\rm opt}} = \max_{\left\{\bff_k\right\}_{k=1}^{K}} ~~ \sum_k \log_2\left(\frac{|\bh_{k}^T\bff_{k}|^2}{\alpha_{k-1}|\bg_{k}^T\bff_{k}|^2 + \frac{1}{\rho_{k-1,{\rm (d)}}}}\right) ~~\rm{s.t.}~~\|\bff_k\|^2 = 1,~~ k = 1,\ldots,K  \label{eqn:C-Optimization4} \ .
\end{align}
The $k^{\rm th}$ base station uses CSI of $\bh_k$ and $\bg_k$ to obtain the optimal $\bff_k$, implying that the sum-rate can be approximately maximized by each base station exchanging CSI only with its neighbors. This eliminates the need for global CSI knowledge and reduces the load on the finite-capacity backhaul link. 

Now, \eqref{eqn:C-Optimization4} deals with maximizing a function consisting of $K$ independent variables, $\{\bff_k\}_{k=1}^K$. Hence, it can be split into $K$ independent problems. Thus, the high SINR approximation is used to remove the interdependency of the users in \eqref{eqn:SumRate}. The optimal linear beamforming vector, $\bff_{k,{\rm opt}}$, is the solution to
\begin{equation}
\bff_{k,{\rm opt}} =  \arg\max_{\bff:\|\bff\|^2 = 1} ~~ \frac{\bff^*\bR_{\bh_k}\bff}{\bff^*\bR_{\bg_k}\bff},   \label{eqn:C-Optimization7}
\end{equation}
where $\bR_{\bh_k} = \bh_k\bh_k^*$ and $\bR_{\bg_k} = \rho_{k-1,{\rm (i)}}\bg_k\bg_k^* + \bI_\Nt$, and $\rho_{k-1,{\rm (i)}} = \alpha_{k-1} \rho_{k-1,{\rm (d)}}$ is the received SNR of the intereference signal at the $(k-1)^{th}$ user. The expression in \eqref{eqn:C-Optimization7} is the well known generalized Rayleigh quotient \cite{Borga1998}. Since $\bR_{\bh_k}$ and $\bR_{\bg_k}$ are Hermitian and $\bR_{\bg_k}$ is positive definite\footnote{$\bR_{\bg_k}$ is positive definite (and hence, full rank) as $\bx^*\bR_{\bg_k}\bx = \bx^*(\rho_{k-1,{\rm (i)}}\bg_k\bg_k^* + \bI_\Nt)\bx > 0$, for all $\bx\neq\mathbf{0}$.}, the solution to \eqref{eqn:C-Optimization7} is given by the generalized eigenvalue decomposition:
\begin{equation}
\bR_{\bh_k}\bff = \lambda_k\bR_{\bg_k}\bff \label{eqn:C-Optimization8},
\end{equation}
where $\lambda_k$ denotes the eigenvalues of $\bR_{\bg_k}^{-1}\bR_{\bh_k}$ \cite{Cadambe2008}. Due to the rank deficient nature of $\bR_{\bh_k}(= \bh_k\bh_k^*)$, there exists only one non-zero eigenvalue. The solution to \eqref{eqn:C-Optimization7} will then be equal to the generalized eigenvector corresponding to the non-zero (maximum) eigenvalue. The solution for $\bff_{k,{\rm{opt}}}$ is invariant to an angular rotation, $\theta$, of the beamforming vector, $\sfee^{j\theta}\bff_{k,{\rm{opt}}}$. Hence, there can be infinitely many solutions to \eqref{eqn:C-Optimization7}. The unit-norm and rotationally invariant nature of $\bff_{k,{\rm{opt}}}$ implies that it is a point on the Grassmann manifold, used popularly in single-cell limited-feedback literature \cite{Love2008}.  

Note that while \eqref{eqn:C-Optimization8} can be used to approximately maximize the sum-rate for both an infinite linear array (in Fig.~\ref{fig:WynerModel}) and a circular arrangement of cells, it can also be used for a finite cellular array by accounting for edge effects and using \eqref{eqn:C-Optimization8} for $k=2,\ldots,K$. The base station in the first cell of the array does not cause interference ($\alpha_0 = 0$), implying that optimal beamforming vector is given by \eqref{eqn:MaxSNR-BFF}. Hence,  \eqref{eqn:C-Optimization8} can be used for any $K$, while the solution in \cite{Lee2008,Lee2009} approximately maximizes sum-rates for only $K=2$. Also, while it is recognized that the ratio in \eqref{eqn:C-Optimization4} is similar to the signal-to-leakage-noise-ratio (SLNR) proposed in \cite{Sadek2007} for multiuser MIMO, the single-cell SLNR solution does not bear a direct impact on the sum-rates obtained and is shown to yield low sum-rates in the high SINR regime \cite{Lim2009}. In contrast, we show in \eqref{eqn:C-Optimization4} that (approximately) maximizing the sum-rate at high SINR is equivalent to maximizing SLNR at each base station for the multicell setup in Section \ref{sec:SysModel}. 


\section{Designing Beamforming Vectors Using Limited Feedback} \label{sec:BFVec_PartialCSI}
In Section~\ref{sec:BFVec_FullCSI}, we presented a beamforming approach to approximately maximize the sum-rate at high SINR in a multicell setting by assuming that full CSI of the desired and interfering channels was available at the base station. A reasonable way to inform the base station about the channel state is through a limited feedback channel. In this section, we propose a limited feedback approach for multicell generalized eigenvector beamforming, where the quantized CSI of both the desired and interfering channels is fed back over the feedback link. The available feedback resources can be utilized efficiently in the multicell case by taking into account the relative strengths of the desired and interfering channels. Note that this is in contrast to single-cell multiuser limited feedback case, where only the CSI of a single channel is fed back \cite{Zhang2009a, Trivellato2008}. 

The limited feedback model for the multicell setup in Section~\ref{sec:SysModel} is described in Fig.~\ref{fig:LFModel}. It is assumed that the $k^{\rm th}$ user can perfectly estimate the desired and interfering channels, $\bh_k$ and $\bg_{k+1}$, using separate training symbols from the $k^{\rm th}$ and  $(k+1)^{\rm th}$ base stations, respectively. The $k^{\rm th}$ receiver uses $\Bd$ and $\Bi$ bits to feed back quantized versions of the $\bh_k$ and $\bg_{k+1}$, respectively, over a limited feedback channel where $\Bd + \Bi = \Bt$ is fixed. Note that $\Bd$ and $\Bi$ will depend on the relative strength of the interfering and desired signals ($\alpha_k$). For example, when $\alpha_k \rightarrow 0$ at the cell center, the contribution of the desired channel towards the SINR is greater than that of the interfering channel and it becomes more important to reduce the loss due to quantization of $\bh_k$ as compared to that due to $\bg_{k+1}$. This can be done by allocating most of the feedback bits to quantize $\bh_k$, i.e. $\Bd \approx \Bt$. Hence, the two channels, $\bh_k$ and $\bg_{k+1}$, are quantized separately using two separate codebooks of (variable) sizes $2^\Bd$ (denoted by $\cW_{k,{\rm (d))}}$) and $2^\Bi$ (given by $\cW_{k,{\rm (i)}}$), respectively.  

The unit-norm direction of the estimated desired (and interfering) channel is given by $\tilde{\bh}_k = \bh_k/\|\bh_k\|$ ($\tilde{\bg}_{k+1} = \bg_{k+1}/\|\bg_{k+1}\|$). The channel directions are then mapped to the respective codebook entries with the smallest angular separations. The quantized vectors for the desired and interfering channels, denoted by $\hat{\bh}_{k}$ and $\hat{\bg}_{k+1}$, respectively, are obtained using
\begin{align}
\hat{\bh}_k =  \arg \max_{\bw \in \cW_{k,{\rm (d)}}} |\tilde{\bh}_k^*\bw|^2 &= \arg\max_{\bw \in \cW_{k,{\rm (d)}}}\cos^2\left(\angle( \tilde{\bh}_k,{\bw})\right) , ~{\rm and}\label{eqn:RVQ-hQuant} \\
\hat{\bg}_{k+1} =  \arg \max_{\bw \in \cW_{k,{\rm (i)}}} |\tilde{\bg}_{k+1}^*\bw|^2 &= \arg\max_{\bw \in \cW_{k,{\rm (i)}}}\cos^2\left(\angle( \tilde{\bg}_{k+1},{\bw})\right) \label{eqn:RVQ-gQuant} .
\end{align}
Since the channel gains, $\|\bh_k\|$ and $\|\bg_{k+1}\|$, are scalars and can be quantized easily, we concentrate on quantizing the channel directions, $\tilde{\bh}_k$ and $\tilde{\bg}_{k+1}$ and assume that the base stations have perfect knowledge of $\|\bh_k\|$. This is a popular assumption made in multiuser MIMO literature \cite{LF-Assumption1, LF-Assumption2, Huang2009}. As quantizing channel gain is (to the best of our knowledge) not investigated for multicell systems, we use the perfect $\|\bh_k\|$ and $\|\bg_{k+1}\|$ knowledge at base station assumption. We will analyze the impact of quantizing channel gain in our future work on limited feedback for multicell systems. 

From \eqref{eqn:C-Optimization4}, it is seen that the $k^{\rm th}$ base station requires information about both $\bh_k$ and $\bg_k$ to compute $\bff_k$. In the limited feedback model presented in this paper, the $k^{\rm th}$ base station has information of both $\hat{\bh}_{k}$ and $\hat{\bg}_{k+1}$ (in addition to perfect knowledge of $\|\bh_k\|$ and $\|\bg_{k+1}\|$). The interfering channel information, $\|\bg_{k+1}\|$, is then sent from the $k^{\rm th}$ to the $(k+1)^{\rm th}$ base station via the backhaul link. In this manner, each base station has knowledge of not only its own desired channel, but also of the interference that is causing to the user in the adjacent cell.

As codebook design for multicell systems is a topic of ongoing research, we use RVQ for channel quantization to facilitate analysis. If we use $B$ bits for feedback, then each of the $2^B$ codebook vectors is independently chosen from the isotropic distribution on the $\Nt$ dimensional unit sphere \cite{Santipach2004, Santipach2005}. The beamforming vector at the $k^{\rm th}$ base station using limited CSI feedback, $\hat{\bff}_k$ is then computed as the generalized eigenvector satisfying
\begin{equation}
\bR_{\hat{\bh}_k}\hat{\bff_k} = \lambda_k\bR_{\hat{\bg}_k}\hat{\bff_k} \label{eqn:EigValDecomp-PartialCSI},
\end{equation}
where $\bR_{\hat{\bh}_k} = \|\bh_k\|^2{\hat{\bh}_k}{\hat{\bh}_k}^*$ and $\bR_{\hat{\bg}_k} = \rho_{k-1,{\rm (i)}}\|\bg_k\|^2\hat{\bg}_k\hat{\bg}_k^* + \bI_\Nt$. Quantization of CSI leads to a loss in the sum-rate. In this paper, we develop a feedback-bit partitioning strategy to approximately minimize the mean loss in sum-rate given by
\begin{align}
\bbE\{\Delta R_s\} \approx & \bbE\{R_{s,{\rm full}} - R_{s,{\rm LF}} \} \nonumber \\
=& \bbE\left\{\sum_k \log_2\left(\frac{|\bh_{k}^T\bff_{k,{\rm full}}|^2}{\alpha_{k}|\bg_{k+1}^T\bff_{k+1,{\rm full}}|^2 + \frac{1}{\rho_{k,{\rm (d)}}}}\right) - \sum_k \log_2\left(\frac{|\bh_{k}^T\hat{\bff}_{k}|^2}{\alpha_{k}|\bg_{k+1}^T\hat{\bff}_{k+1}|^2 + \frac{1}{\rho_{k,{\rm (d)}}}}\right) \right\} \label{eqn:DeltaR_main} ,
\end{align}
where $R_{s,{\rm full}} $ is the sum-rate obtained using the high SINR approximation with full CSI at the base station, as given in  \eqref{eqn:C-Optimization4} and $R_{s,{\rm LF}}$ refers to the sum-rate using limited feedback.  Also, $\bff_{k+1,{\rm full}}$ corresponds to the full CSI beamforming vector (from \eqref{eqn:C-Optimization8}). 


\section{Optimizing Feedback Bits to Minimize the Mean Loss in Sum-Rate at High SINR} \label{sec:OptFB_PartialCSI}
The beamforming strategy proposed in Sections~\ref{sec:SINRMax} and \ref{sec:BFVec_PartialCSI} requires  channel state information of both the desired and interfering channel at the base station. As the number of available feedback bits, $\Bt$ is fixed, it is possible that allocating feedback bits between the desired and interfering channels can further improve the limited feedback sum-rate. In this section, we propose a feedback-bit allocation strategy to numerically evaluate the number of bits required to quantize the desired and interfering channels using RVQ. We model all the channels by the Rayleigh fading model, where each entry is a zero-mean unit-variance complex Gaussian independent and identically distributed (i.i.d.) random variable according to $\cN_c(0,1)$. While it is recognized that Rayleigh fading does not model realistic propagation channels accurately, we use the i.i.d. assumption in the limited feedback analysis to obtain closed-form expressions of the mean loss in sum-rate and feedback-bit partitioning algorithms. 

Maximizing the mean sum-rate using generalized eigenvector beamforming with limited feedback is equivalent to minimizing $\bbE\{\Delta R_s\}$ in \eqref{eqn:DeltaR_main}, which is rewritten as
\begin{equation}\label{eqn:DeltaR2}
\bbE\{\Delta R_s\} = \sum_k\underbrace{\bbE\left\{\log_2\left(\frac{|\bh_k^T\bff_{k, {\rm full}}|^2}{|\bh_k^T\hat{\bff_k}|^2} \right)\right\}}_{\Td} + \sum_k \underbrace{\bbE\left\{\log_2\left(\frac{\rho_{k, {\rm (i)}} |\bg_{k+1}^T\hat{\bff}_{k+1}|^2 + 1}{\rho_{k, {\rm (i)}}|\bg_{k+1}^T\bff_{k+1,{\rm full}}|^2 + 1}\right) \right\}}_{\Ti} \ , 
\end{equation}
by interchanging the terms inside the logarithms. The mean loss in sum-rate can be viewed to be a contribution of two terms, namely $\Td$ and $\Ti$, corresponding to the mean loss resulting from quantizing the desired and interfering channels, respectively. Note that $\Td$ and $\Ti$ depend only on $\Bd$ and $\Bi$, the number of bits used to quantize the desired and intereference channels, respectively, and are independent of $T_{l,{\rm (d)}}$ and $T_{l,{\rm (i)}}$, for $k \neq l$. Therefore, minimizing \eqref{eqn:DeltaR2} is equivalent to minimizing $\Delta_k  = \Td + \Ti$ for each $k$. Since obtaining a closed form expression for \eqref{eqn:DeltaR2} is complicated, we derive an upper bound for $\bbE\{\Delta R_s\}$ in terms of bounds on $\Td$ and $\Ti$. 

Before we proceed, we present a lemma that is used to derive an upper bound on \eqref{eqn:DeltaR2}. Let $\cH$ denote the set of all unit-norm channel directions, $\tilde{\bh}$.
 
\begin{lemma}\label{lem:meanLogCos}
The mean of $\log_2\left(\cos^2\left(\angle(\tilde{\bh},\hat{\bh})\right)\right)$ (in \eqref{eqn:RVQ-hQuant}) is given by
$$
\bbE_{\cH,\cW} \left\{\log_2 \left(\cos^2\left(\angle(\tilde{\bh},\hat{\bh})\right)\right)\right\} = \log_2(e)\sum_{i=0}^{2^B}{{2^B}\choose i}(-1)^{i} \sum_{l=1}^{i(\Nt-1)}\frac{1}{l} \ .
$$
\end{lemma}
\begin{proof}
The proof is given in Appendix \ref{app:meanLogCos}.
\end{proof}
The following theorems present upper bounds on $\Td$ and $\Ti$.

\begin{theorem}\label{thm:Tdes-UppBnd}
The mean loss in sum-rate due to quantizing desired channel of the $k^{\rm th}$ user, $\Td$ in \eqref{eqn:DeltaR2} is upper-bounded by
$$
 \Td \leq -\log_2(e)\sum_{i=0}^{2^\Bd}{{2^\Bd}\choose i}(-1)^{i} \sum_{l=1}^{i(\Nt-1)}\frac{1}{l} \ .
$$
\end{theorem}
\begin{proof}
The proof is given in Appendix \ref{app:Tdes-UppBnd}.
\end{proof}

\begin{theorem}\label{thm:Tint-UppBnd}
The mean loss in sum-rate due to quantizing interfering channel of the $k^{\rm th}$ user, $\Ti$ in \eqref{eqn:DeltaR2} is upper-bounded by
$$
 \Ti \leq \log_2\left(1 + \rho_{k,{\rm (i)}}\Nt 2^\Bi\beta\left(2^\Bi , \frac{\Nt}{\Nt-1}\right) \right) \ .
$$
\end{theorem}
\begin{proof}
The proof is given in Appendix \ref{app:Tint-UppBnd}.
\end{proof}

Using Theorems~\ref{thm:Tdes-UppBnd} and \ref{thm:Tint-UppBnd}, the difference in the mean sum-rates assuming full-CSI and using limited feedback, $\bbE\{\Delta R_s\}$, is upper bounded by
\begin{eqnarray}
\bbE\{\Delta R_s\} & \leq & \sum_k \log_2\left(1 + \rho_{k,{\rm (i)}}\Nt 2^\Bi\beta\left(2^\Bi, \frac{\Nt}{\Nt-1}\right) \right) \nonumber \\
& - &\sum_k \log_2(e)\sum_{i=0}^{2^\Bd}{{2^\Bd}\choose i}(-1)^{i} \sum_{l=1}^{i(\Nt-1)}\frac{1}{l} 
\label{eqn:finalDelta_Kcells}\ .
\end{eqnarray}
For a system with two transmit antennas at each base station, we substitute $\Nt =2$ in \eqref{eqn:finalDelta_Kcells} to get
\begin{equation}
\bbE\{\Delta R_s\} \leq \sum_k \log_2\left(1 + 2\rho_{k,{\rm (i)}} 2^\Bi\beta\left(2^\Bi, 2\right) \right)  -\sum_k\log_2(e)\sum_{i=0}^{2^\Bd}{{2^\Bd}\choose i}(-1)^{i} \sum_{l=1}^{i}\frac{1}{l} \label{eqn:finalDelta_2cells1} \ .
\end{equation}
Using series expansions, it can be shown that
$\sum_{i=0}^{2^\Bd}{{2^\Bd}\choose i}(-1)^{i}
\sum_{l=1}^{i}\frac{1}{l} = -2^{-\Bd}$. Further, $\beta\left(2^\Bi,
2\right) = 1/(2^\Bi(2^\Bi + 1))$ and $\Bi = \Bt - \Bd$. Hence, \eqref{eqn:finalDelta_2cells1} is rewritten as
\begin{equation}
\bbE\{\Delta R_s\} \leq \sum_k \underbrace{\log_2\left(1 + 2\rho_{k, {\rm (i)}} \frac{1}{2^{\Bt -\Bd} + 1}\right) + 2^{-\Bd}\log_2(e)}_{\tilde{\Delta}_k}  \label{eqn:finalDelta_2cells2}\ .
\end{equation}
To find the optimum number of bits to quantize the desired and interfering channels, we use the relation $\Bi = \Bt - \Bd$ in \eqref{eqn:finalDelta_2cells2}. For simplicity, we denote the right hand side of expression \eqref{eqn:finalDelta_2cells2} by $\tilde{\Delta}_k$. Treating $\Bd$ as a real number, we first show in Theorem~\ref{thm:Delta_Convexity} that $\tilde{\Delta}_k$ is convex in $\Bd \in [0,\Bt]$, and then use the
result to compute the optimum number of desired and interfering feedback bits.

\begin{theorem}\label{thm:Delta_Convexity}
The minimum value of the upper bound, $\tilde{\Delta}_k$ in \eqref{eqn:finalDelta_2cells2}, is obtained at the value of $\Bd$ equal to
$$
\Bd^{\rm real} = \Bt - \log_2\left(1+\rho_{k, {\rm (i)}} + \sqrt{\rho_{k, {\rm (i)}} 2^{\Bt+1} + (\rho_{k, {\rm (i)}})^2}\right),
$$
where $\Bd^{\rm real}  \in [0,\Bt]$ is a real number.
\end{theorem}
\begin{proof}
The proof is given in Appendix \ref{app:Delta_Convexity}.
\end{proof}

Note that the minimization in Theorem~\ref{thm:Delta_Convexity} is over the set of real values and, hence, $\Bd^{\rm real}$ is not necessarily an interger. Since the upper bound $\tilde{\Delta}_k$ is convex in $\Bd^{\rm real}$, we only need to consider the ceiling and floor of $\Bd^{\rm real}$ (denoted by $\lfloor \Bd^{\rm real} \rfloor$ and $\lceil \Bd^{\rm real} \rceil$, respectively) to find the optimal number of bits to quantize the desired channel \cite{Gokbayrak1999,Murota2003}, as given in Corollary~\ref{Cor:OptBd}.

\begin{corollary} \label{Cor:OptBd}
The optimum number of desired and interfering feedback bits are given by $\Bd^{\rm opt}$ and $\Bi^{\rm opt} = \Bt - \Bd^{\rm opt}$ respectively, where $\Bd^{\rm opt}$ is either $\lfloor \Bd^{\rm real} \rfloor$ or $\lceil \Bd^{\rm real} \rceil$. 
\end{corollary}

Since $\rho_{k, {\rm (i)}}=\rho_{k, {\rm (d)}}\alpha_k$, it is clear from Theorem~\ref{thm:Delta_Convexity} that $\Bd^{\rm real}$ (and hence, $\Bd^{\rm opt}$) is a function of $\rho_{k, {\rm (d)}}$, $\alpha_k$, and $\Bt$. Note that for $\alpha_k = 0$, $\Bd^{\rm opt} = \Bt$. This implies that when the interfering signal has zero power (or there is no interferer transmitting), all the available bits will be used to quantize the desired signal. In the absence of interference, it is evident that the strength of the received signal, $\rho_{k, {\rm (d)}}$ does not affect the bits allocated. In contrast, when $\alpha_k = 1$, $\Bd^{\rm opt}$ increases as $\rho_{k, {\rm (d)}}$ decreases. This makes intuitive sense because if the desired signal strength is high, fewer feedback bits are assigned to the desired channel and vice versa. In Section~\ref{sec:Results}, we show that the values of ($\Bd,\Bi$) from Theorem~\ref{thm:Delta_Convexity} match the simulation results.


\section{Simulation Results} \label{sec:Results}
In this section, we present simulation results for the case when full CSI is available at the base station and then present results for the limited feedback scenario. We show that the feedback-bit allocation strategy described in Section~\ref{sec:OptFB_PartialCSI} approximately minimizes the mean loss in sum-rate due to quantization. Simulations are also used to verify that the partitioning of feedback bits between the desired and interfering channels proposed in Theorem~\ref{thm:Delta_Convexity} and Corollary~\ref{Cor:OptBd} matches numerical results.

Unless otherwise stated, we assume that all $K$ users have the same received desired and interfering signal strengths, i.e., $\rho_{k,{\rm (d)}} = \rho_{\rm (d)}$  and $\alpha_{k} = \alpha$, for all $k$, for simplicity. We provide simulation results for the asymmetric case at the end of this section. 

\subsection{Full CSI Case}\label{sec:Results-FullCSI}
In Section~\ref{sec:BFVec_FullCSI}, a high SINR approximation was made to approximately maximize the sum-rates in a distributed fashion across base stations. In Fig.~\ref{fig:FullCSI_HighSINRApprox},  we verify that the approximation yields sum-rates that are reasonably close to the actual values for $\rho_{\rm (d)}$ values as small as $5~dB$, for $\Nt = K = 4$ and $\alpha = \{0.001,0.1,1\}$. In Fig.~\ref{fig:SumRate_GEBF_DPC_IncrK}, we show that as $K$ (number of cells or users) increases, GEBF yields sum-rates that are very close to the upper-bound using multicell DPC for $K$ users for $\Nt = 4$ and $\rho_{\rm (d)} = 10~dB$. Though the difference in sum-rates is maximum for $\alpha = 1$ because of a general drop in the SINR of each user due to increased interference from adjacent cells, it is seen that the multicell GEBF presented in this paper yields sum-rates reasonably close to the upper bound from DPC. 



\subsection{Limited CSI Feedback}\label{sec:Results-LF}
The upper bound on the reduction in sum-rate (due to quantization) was derived in Section~\ref{sec:OptFB_PartialCSI}, by assuming that the number of bits available for both, the desired and interfering channels are large enough to ignore quantization errors. Hence, in the simulation results presented in this section, for a $\Bt$ bit feedback system, we assume that $\Bd \in [3,\Bt-3]$, as using RVQ codebooks of size $2^3 = 8$ leads to low values of quantization error. In Fig.~\ref{fig:UppBndComp}, for $\rho_{\rm (d)} = 10~dB$, $\Bt = 15$, $K=\Nt = 2$ and $\alpha = \{0.001,0.1,1\}$, it is shown that the upper bound derived in \eqref{eqn:finalDelta_2cells2} is reasonably tight near the minimum mean loss in sum-rate.  It is evident that partitioning adaptively the feedback bits, as a function of the location of the mobile terminal, is an effective method to increase the sum-rates obtained in limited feedback systems. This implies that the upper bound derived in \eqref{eqn:finalDelta_2cells2} can be used to evaluate the number of quantization bits needed to minimize the mean sum-rate loss. Further, Fig.~\ref{fig:UppBndComp} clearly demonstrates the convexity of the actual loss in sum-rate and the upper bound in \eqref{eqn:finalDelta_2cells2}.

Next, we present simulation results to show that GEBF with the proposed feedback-bit partitioning algorithm outperforms non-cooperative eigen-beamforming and multicell zero-forcing beamforming strategies with limited CSI feedback. Since the eigen-beamforming approach described in Section~\ref{sec:BFVec_PartialCSI} is non-cooperative, all the available feedback bits are used to quantize the desired channel only, i.e. $\Bd^{\rm EBF} = \Bt$, whereas for multicell zero-forcing approach, $\Bd^{\rm ZF}  = \Bi^{\rm ZF} = \Bt/2$ is assumed. Fig.~\ref{fig:CompTxStrat} shows the sum-rates that can be obtained using the three transmission strategies for a two-cell (two-user) scenario with $\Bt = 6$ and $\rho_{\rm (d)} = 10~dB$ and $K=\Nt = 2$. It is seen that the proposed limited feedback algorithm yields higher sum-rates. As $\alpha$ increases, the sum-rates using both GEBF and non-cooperative eigen-beamforming reduce, but that of eigen-beamforming falls off more drastically. This is because as $\alpha \rightarrow 1$, interference becomes significant and the beamforming vector needs to reduce the interference in addition to increasing the desired signal strength. Zero forcing with full CSI yields a constant sum-rate for all values of $\alpha$ due to the nulling out of the interference term in the denominator of \eqref{eqn:SINR2}.

The sum-rates obtained using the proposed bit-partitioning technique as $\Bt$ increases are compared in Fig.~\ref{fig:SumRate_Vs_Btot_3Strat} for $\Nt = K = 2$ and $\rho_{\rm (d)} = 10~dB$ for $\alpha = \{0.001,0.1,1\}$. It is seen that the cell-edge users ($\alpha \rightarrow 1$) requires a larger $\Bt$ as compared to users in the cell-center ($\alpha \rightarrow 1$). This is due to the requirement of cell-edge users to quantize the interfering channel (in addition to the desired channel) with a sufficiently high resolution. In contrast, users in the cell-center that have weak interfering signals have to only ensure that the desired channel is allocated sufficient number of feedback bits. Hence, the feedback-bits, $\Bt$ can be varied adaptively as a function of the user location within the cell to yield a given sum-rate. 

In Fig.~\ref{fig:BdBi_Vs_LogAlpha}, we show the variation in the number of bits allocated to the desired and interfering channels as a function of the interfering to desired signal ratio, given by $\tilde{\alpha} = 10\log_{10}(\alpha)~dB$. In Fig.~\ref{fig:BdBi_Vs_LogAlpha}, for $K = \Nt = 2$ and $\rho_{\rm (d)} = 10~dB$, we can see that when the difference in path-loss is greater than $37~dB$, $(\Bd,\Bi) = (8,0)$. As the path-loss difference decreases, $\Bd$ reduces and $\Bi$ increases, due to an increasing need to quantize the interfering channel with greater resolution. When both the desired and interfering channels have the same signal strength, it is seen that $(\Bd,\Bi) = (2,6)$, i.e. reducing the interfering signal strength is `more important' to maximize the sum-rate than increasing the desired signal strength. 

Finally, we consider the case where the users are randomly located in their cells and have different interfering to desired signal ratios, i.e. $\tilde{\alpha}_k \neq \tilde{\alpha}_l, k\neq l$. Hence, the users can have different feedback bit allocations. The performance of the proposed feedback-bit allocation strategy with GEBF can be examined for the asymmetric scenario by evaluating the average data-rate in a cell, where the average is taken over all possible user locations in the cell. Assuming $\Bt = 6$ at each user, we consider $1000$ cells (users) arranged in a linear fashion, with $\tilde{\alpha}_k$ uniformly distributed between $[-40~dB, 0~dB]$, for all $k$. Adjacent cells cooperate by sharing the interfering channel CSI to compute the bit partitions for each user independently.  In Fig.~\ref{fig:Bits6_diffAlphas}, it is seen that the average data-rate in a cell using the proposed feedback-bit allocation strategy is reasonably close to that obtained using GEBF with full CSI. We also show that the data rate obtained using equal bit partitioning in GEBF, i.e. $\Bd = \Bi = \Bt/2$ for all $k$ is less than that obtained using the proposed bit allocation, especially for larger $\rho_{\rm (d)}$. This makes intuitive sense as the equal bit allocation does not reduce the quantization error, which grows with larger $\rho_{\rm (d)}$, while the proposed strategy allocates bit adaptively to minimize the overall impact of quantization error.  We also plot the average data rates obtained using the non-cooperative eigenbeamforming and multicell zero-forcing approaches (described in Section~\ref{sec:ExistingStrategies-FullCSI}) with limited feedback. The proposed feedback-bit strategy with GEBF outperforms both, EBF and ZF, especially at larger values of $\rho_{\rm(d)}$. Hence, it can be concluded that the proposed limited feedback strategy is effective in multicell systems.

\section{Conclusion} \label{sec:Conclusion}
We first presented a beamforming strategy that uses cooperation among base stations in a multicell MISO system to approximately maximize the sum-rate at high SINR. A linear array of cells based on the Wyner model is considered, in addition to a circular extension of the model. For the limited feedback scenario, we presented a feedback-bit allocation strategy to approximately minimize the mean loss in sum-rate caused due to quantization using RVQ. The proposed technique relies on the relative strength between the desired and interfering signals within the cell to allocate bits adaptively to each of the two channels. A closed form expression for the number of feedback bits was derived.  Using simulations, it was shown that the multicell beamforming approach yields sum-rates reasonably close to those obtained using multicell DPC, with full CSI. The average data rate in a cell using the feedback-bit allocation strategy was also shown to reasonably close to the full CSI case, verifying that the proposed multicell limited feedback algorithm yields high sum-rates using partial cooperation.

\appendices


\section{Proof of Lemma~\ref{lem:meanLogCos}}\label{app:meanLogCos}
It was shown in \cite{Au-Yeung2007} that the probability density function of $\nu = \cos^2\left(\angle(\tilde{\bh},\hat{\bh})\right)$ is given by
$$
f_\nu(\nu) = \sum_{i=0}^{N}\sum_{j=1}^{i(\Nt-1)} {N \choose i} {{i(\Nt-1)}\choose j}(-1)^{i+j} j \nu^{j-1},
$$
where $N = 2^\Bi$ is the size of the RVQ codebook. The mean, $\bbE\{\log_2(\nu)\}$, is then computed as
\begin{equation}
\bbE\{\log_2(\nu)\} = \log_2(e)\int_0^1\ln(\nu) f_\nu(\nu) d\nu \label{eqn:pdf-Logcos0},
\end{equation}
which is written as
\begin{eqnarray}
\bbE\{\log_2(\nu)\} &=&  \log_2(e)\sum_{i=0}^{N}{N \choose i} (-1)^i \sum_{j=1}^{i(\Nt-1)} {{i(\Nt-1)}\choose j}(-1)^{j} j \int_0^1\ln(\nu)  \nu^{j-1} d\nu \label{eqn:pdf-Logcos1} \\
 & = & \log_2(e)\sum_{i=0}^{N}{N \choose i} (-1)^i \sum_{j=1}^{i(\Nt-1)} {{i(\Nt-1)}\choose j} \frac{(-1)^{j+1}}{j^2} \label{eqn:pdf-Logcos2} \\
 & = & \log_2(e)\sum_{i=0}^{N}{N \choose i} (-1)^i \sum_{l=1}^{i(\Nt-1)} \frac{1}{l} \label{eqn:pdf-Logcos3}
\end{eqnarray}
where substituting $f_\nu(\nu)$ in \eqref{eqn:pdf-Logcos0} leads to \eqref{eqn:pdf-Logcos1}. Also, \eqref{eqn:pdf-Logcos2} is obtained by solving for $\int_0^1\ln(\nu)  \nu^{j-1} d\nu$ in \eqref{eqn:pdf-Logcos1}. Finally, \eqref{eqn:pdf-Logcos3} is given in \cite{intSeriesBook}.

\section{Proof of Theorem~\ref{thm:Tdes-UppBnd}}\label{app:Tdes-UppBnd}
Dropping the index $k$ for the sake of convenience, the first term of \eqref{eqn:DeltaR2}, $\Td$ is rewritten in terms of the channel directions, $\tilde{\bh}$ as
\begin{equation}
\bbE\left\{\log_2\left(\frac{|\bh^T\bff_{\rm full}|^2}{|\bh^T\hat{\bff}|^2} \right)\right\} = \bbE\left\{\log_2(|\tilde{\bh}^T\bff_{\rm full}|^2) \right\} - \bbE\left\{\log_2(|\tilde{\bh}^T\hat{\bff}|^2) \right\} \label{eqn:Tdes-UppBnd1} \ .
\end{equation}
Now, since $|\bh^T\bff_{\rm full}| \in [0,1]$, and $|\bh^T\hat{\bff}| \in [0,1]$, we can safely assume that the angles $ \angle(\tilde{\bh}^c,\hat{\bff}),  \angle(\hat{\bh}^c,\hat{\bff}) \in [0,\pi/2]$. Using the triangle inequality for angles \cite{IntTheoryBook}, we write
\begin{eqnarray}
\angle(\tilde{\bh}^c,\hat{\bff}) &\leq& \angle(\tilde{\bh}^c,\hat{\bh}^c) + \angle(\hat{\bh}^c,\hat{\bff}), \nonumber\\
& = & \angle(\tilde{\bh},\hat{\bh}) + \angle(\hat{\bh}^c,\hat{\bff}) \label{eqn:triangleIneq},
\end{eqnarray}
where $\angle(\tilde{\bh},\hat{\bh})=\angle(\tilde{\bh}^c,\hat{\bh}^c)$ is used to obtain \eqref{eqn:triangleIneq}. We can lower bound $|\tilde{\bh}^T\hat{\bff}|^2 = \cos^2(\angle(\tilde{\bh}^c,\hat{\bff})) $ by
\begin{eqnarray}
\cos^2(\angle(\tilde{\bh}^c,\hat{\bff})) &\geq& \left[\cos(\angle(\tilde{\bh},\hat{\bh}))\cos(\angle(\hat{\bh}^c,\hat{\bff})) - \sin(\angle(\tilde{\bh},\hat{\bh})) \sin (\angle(\hat{\bh}^c,\hat{\bff})) \right]^2 \label{eqn:Tdes-UppBnd1a}\\
&\approx &  \left[\cos(\angle(\tilde{\bh},\hat{\bh}))\cos(\angle(\hat{\bh}^c,\hat{\bff}))\right]^2 \label{eqn:Tdes-UppBnd2} \\
& = & \cos^2(\angle(\tilde{\bh},\hat{\bh}))|\hat{\bh}^T\hat{\bff}|^2 \label{eqn:Tdes-UppBnd3}
\end{eqnarray}
where \eqref{eqn:Tdes-UppBnd2} is obtained by understanding that as the number of quantization bits increase, $\angle(\tilde{\bh},\hat{\bh})$ will reduce and hence, $\sin(\angle(\tilde{\bh},\hat{\bh}))$ will be very small. The product $\sin(\angle(\tilde{\bh},\hat{\bh})) \sin (\angle(\hat{\bh},\hat{\bff}))$ can, therefore, be ignored.

Substituting \eqref{eqn:Tdes-UppBnd3} in \eqref{eqn:Tdes-UppBnd1}, we get
\begin{eqnarray}
\bbE\left\{\log_2\left(\frac{|\bh^T\bff_{\rm full}|^2}{|\bh^T\hat{\bff}|^2} \right)\right\} &\leq& \bbE\left\{\log_2(|\tilde{\bh}^T\bff_{\rm full}|^2) \right\}  -  \bbE\left\{\log_2(\cos^2(\angle(\tilde{\bh},\hat{\bh}))|\hat{\bh}^T\hat{\bff}|^2 ) \right\} \nonumber \\
& = & -  \bbE\left\{\log_2(\cos^2(\angle(\tilde{\bh},\hat{\bh}))\right\} \label{eqn:Tdes-UppBnd4} \\
& = &  -  \log_2(e)\sum_{i=0}^{2^B}{{2^\Bd}\choose i}(-1)^{i} \sum_{l=1}^{i(\Nt-1)}\frac{1}{l} \label{eqn:Tdes-UppBnd5}
\end{eqnarray}
where \eqref{eqn:Tdes-UppBnd4} was obtained using the fact that $\bbE\left\{\log_2(|\tilde{\bh}^T\bff_{\rm full}|^2) \right\}  = \bbE\left\{\log_2(| \hat{\bh}^T\hat{\bff}|^2) \right\}$ and \eqref{eqn:Tdes-UppBnd5}, from Lemma~\ref{lem:meanLogCos}.

\section{Proof of Theorem ~\ref{thm:Tint-UppBnd}}\label{app:Tint-UppBnd}
Dropping the index $k$ (and $(k-1)$) for the sake of convenience, the second term of \eqref{eqn:DeltaR2}, $\Ti$, is upper bounded by
\begin{eqnarray}
\bbE\left\{\log_2\left(\frac{\rho_{\rm (d)}\alpha|\bg^T\hat{\bff}|^2 + 1}{\rho_{\rm (d)}\alpha|\bg^T\bff_{\rm full}|^2 + 1}\right) \right\}   & \leq & \bbE\left\{\log_2(\rho_{\rm (d)}\alpha|\bg^T\hat{\bff}|^2 + 1) \right\} \nonumber \\ 
& \leq & \log_2(1 + \rho_{\rm (d)}\alpha\bbE\{|\bg^T\hat{\bff}|^2\})  \label{eqn:Tint-UppBnd2} \\
& = & \log_2(1 + \rho_{\rm (d)}\alpha\bbE\{\|\bg\|^2\} \bbE\{|\tilde{\bg}^T\hat{\bff}|^2\})  \label{eqn:Tint-UppBnd3}   \\
& = & \log_2(1 + \rho_{\rm (d)}\alpha\Nt \bbE\{|\tilde{\bg}^T\hat{\bff}|^2\})  \label{eqn:Tint-UppBnd4} \ .
\end{eqnarray}
Here, \eqref{eqn:Tint-UppBnd2} is obtained from Jensen's inequality. In \eqref{eqn:Tint-UppBnd3} and \eqref{eqn:Tint-UppBnd4}, we use the relations $\bg = \|\bg\|^2\tilde{\bg}$ and $\bbE\{\|\bg\|^2\} = \Nt$, respectively.

Now, the triangle inequality for angles is rewritten as \cite{IntTheoryBook},
\begin{eqnarray}
\angle(\tilde{\bg}^c,\hat{\bff}) &\geq&  |\angle(\tilde{\bg}^c,\hat{\bg}^c)-\angle(\hat{\bg}^c,\hat{\bff})| \nonumber, \\
 \ &=&  |\angle(\tilde{\bg},\hat{\bg})-\angle(\hat{\bg}^c,\hat{\bff})|  \nonumber, 
\end{eqnarray}
since the angles are all positive and lie between $[0,\pi/2]$.

Using trigonometry, $\sin^2(\angle(\tilde{\bg}^c,\hat{\bff}))$ is lower-bounded by
\begin{eqnarray}
\sin^2(\angle(\tilde{\bg}^c,\hat{\bff})) & \geq & \left[\sin(\angle(\tilde{\bg},\hat{\bg}))\cos(\angle(\hat{\bg}^c,\hat{\bff})) - \cos(\angle(\tilde{\bg},\hat{\bg}))\sin(\angle(\hat{\bg}^c,\hat{\bff}))\right]^2 \label{eqn:Tint-UppBnd5} \\
& \approx & \left[-\cos(\angle(\tilde{\bg},\hat{\bg}))\sin(\angle(\hat{\bg}^c,\hat{\bff}))\right]^2 \label{eqn:Tint-UppBnd6} \\
& \approx & \cos^2(\angle(\tilde{\bg},\hat{\bg}))\label{eqn:Tint-UppBnd7}
\end{eqnarray}
where \eqref{eqn:Tint-UppBnd5} is obtained from the relations $\sin^2(|\theta|) = \sin^2(\theta)$ and $\sin(A-B) = \sin(A)\cos(B) - \sin(B)\cos(A)$. Also, when sufficient quantization bits are used, $\sin(\angle(\tilde{\bg},\hat{\bg}))$ is extremely small. Hence, the product $\sin(\angle(\tilde{\bg},\hat{\bg}))\cos(\angle(\hat{\bg}^c,\hat{\bff}))$ can be ignored when the number of bits used for quantization is sufficiently large. Note that as $\cos(\angle(\hat{\bg}^c,\hat{\bff}))$ is close to 0 (due to the generalized eigenvector relation that minimizes the value of $|\hat{\bg}^T\hat{\bff}|$), $\sin^2(\angle(\hat{\bg}^c,\hat{\bff}))$ is very close to 1. This leads to \eqref{eqn:Tint-UppBnd7} .

We know that $|\tilde{\bg}^T\hat{\bff}|^2 = \cos^2(\angle(\tilde{\bg}^c,\hat{\bff}))$. Using \eqref{eqn:Tint-UppBnd7}, we have
\begin{eqnarray}
\bbE\{|\tilde{\bg}^T\hat{\bff}|^2\} &\leq& \bbE\{1 - \cos^2(\angle(\tilde{\bg},\hat{\bg}))\} \nonumber \\
&=& 2^B\beta\left(2^B, \frac{\Nt}{\Nt-1}\right) \label{eqn:Tint-UppBnd8}
\end{eqnarray}
where \eqref{eqn:Tint-UppBnd8} is obtained from\cite{Au-Yeung2007}.

\section{Proof of Theorem ~\ref{thm:Delta_Convexity}}\label{app:Delta_Convexity}
We denote (from \eqref{eqn:finalDelta_2cells2})
\begin{equation}
\tilde{\Delta}_k = \log_2\left(1 + 2\rho_{k,{\rm (i)}} \frac{1}{2^{\Bt -\Bd} + 1}\right) + 2^{-\Bd}\log_2(e) \label{eqn:finalDelta_2cells3}\ .
\end{equation}
Note that $\tilde{\Delta}_k$ is continuous and differentiable in $\Bd$. The partial derivative of $\tilde{\Delta}_k$ in terms of $\Bd$ is obtained as
\begin{equation}
\frac{\partial\tilde{\Delta}_k}{\partial\Bd} = 2\rho_{k, {\rm (i)}} \frac{2^{\Bt-\Bd}}{\left(2^{\Bt-\Bd}+2\rho_{k,{\rm (i)}}+1\right)\left(1+2^{\Bt-\Bd}\right)}  - 2^{- \Bd}  \label{eqn:finalDelta_2cells4}\ .
\end{equation}
Now, $\tilde{\Delta}_k$ is convex on the convex set $\bbC^{2\times1}$ iff its gradient ${\partial\tilde{\Delta}_k}/{\partial\Bd}$ is monotone, i.e.
$$
\left(\frac{\partial\tilde{\Delta}_k}{\partial\Bd^{(1)}} - \frac{\partial\tilde{\Delta}_k}{\partial\Bd^{(2)}})\right)\left(\Bd^{(1)} - \Bd^{(2)}\right) \geq 0 \ .
$$
From \eqref{eqn:finalDelta_2cells3}, it is shown using simple algebra that if $\Bd^{(1)} < \Bd^{(2)}$, then
\begin{align}
\begin{split}
-2^{- \Bd^{(1)}} &+ 2\rho_{k, {\rm (i)}} \frac{2^{\Bt-\Bd^{(1)}}}{\left(2^{\Bt-\Bd^{(1)}}+2\rho_{k, {\rm (i)}}+1\right)\left(1+2^{\Bt-\Bd^{(1)}}\right)} < \\
&-2^{- \Bd^{(2)}} + 2\rho_{k, {\rm (i)}} \frac{2^{\Bt-\Bd^{(2)}}}{\left(2^{\Bt-\Bd^{(2)}}+2\rho_{k,{\rm (i)}}+1\right)\left(1+2^{\Bt-\Bd^{(2)}}\right)},
\end{split}
\end{align}
implying that the condition for convexity stated in this section is
satisfied and $\tilde{\Delta}_k$ is convex in $\Bd$. Hence, the value
of $\Bd$ that minimizes $\tilde{\Delta}_k$ will be a global optimal
value and is obtained by equating \eqref{eqn:finalDelta_2cells4} to
zero. A closed-form expression for the optimal $\Bd$ is given by
\begin{equation}
\Bd = \Bt - \log_2\left(1+\rho_{k,{\rm (i)}} + \sqrt{\rho_{k,{\rm (i)}}2^{\Bt+1} + (\rho_{k,{\rm (i)}})^2}\right).
\end{equation}

\bibliographystyle{IEEEtran}
\bibliography{NetworkMIMO_Refs_v2}

\begin{thebibliography}{10}
\providecommand{\url}[1]{#1}
\csname url@samestyle\endcsname
\providecommand{\newblock}{\relax}
\providecommand{\bibinfo}[2]{#2}
\providecommand{\BIBentrySTDinterwordspacing}{\spaceskip=0pt\relax}
\providecommand{\BIBentryALTinterwordstretchfactor}{4}
\providecommand{\BIBentryALTinterwordspacing}{\spaceskip=\fontdimen2\font plus
\BIBentryALTinterwordstretchfactor\fontdimen3\font minus
  \fontdimen4\font\relax}
\providecommand{\BIBforeignlanguage}[2]{{%
\expandafter\ifx\csname l@#1\endcsname\relax
\typeout{** WARNING: IEEEtran.bst: No hyphenation pattern has been}%
\typeout{** loaded for the language `#1'. Using the pattern for}%
\typeout{** the default language instead.}%
\else
\language=\csname l@#1\endcsname
\fi
#2}}
\providecommand{\BIBdecl}{\relax}
\BIBdecl

\bibitem{Simeone2009}
O.~Simeone, O.~Somekh, H.~V. Poor, and S.~S. (Shitz), ``Local base station
  cooperation via finite-capacity links for the uplink of wireless networks,''
  \emph{{IEEE} Transactions on Information Theory}, vol.~55, no.~1, pp.
  190--204, Jan. 2009.

\bibitem{Ali2009}
S.~H. Ali and V.~C.~M. Leung, ``Dynamic frequency allocation in fractional
  frequency reused {OFDMA} networks,'' \emph{{IEEE} Transactions on Wireless
  Communications}, vol.~8, no.~8, pp. 4286--4295, Aug. 2009.

\bibitem{Choi2008}
W.~Choi and J.~G. Andrews, ``The capacity gain from intercell scheduling in
  multi-antenna systems,'' \emph{{IEEE} Transactions on Wireless
  Communications}, vol.~7, no.~2, pp. 714--725, Feb. 2008.

\bibitem{Shamai2001a}
S.~Shamai and B.~M. Zaidel, ``Enhancing the cellular downlink capacity via
  co-processing at the transmitting end,'' in \emph{Proc. of {IEEE} Vehicular
  Technology Conference}, May 2001, pp. 1745--1749.

\bibitem{Zhang2004}
H.~Zhang and H.~Dai, ``Cochannel interference mitigation and cooperative
  processing in downlink multicell multiuser {MIMO} networks,'' \emph{{EURASIP}
  Journal on Wireless Communications and Networking}, vol. 2004, no.~2, pp.
  222--235, 2004.

\bibitem{Jafar2004}
S.~A. Jafar, G.~J. Foschini, and A.~J. Goldsmith, ``Phantom{N}et: {E}xploring
  optimal multicellular multiple antenna systems,'' \emph{{EURASIP} Journal of
  Applied Signal Proccessing}, vol. 2004, pp. 591 -- 604, 2004.

\bibitem{Kang2006}
M.~Kang, L.~Yang, and M.-S. Alouini, ``Capacity of {MIMO} channels in the
  presence of co-channel interference,'' \emph{Wireless Communications and
  Mobile Computing}, vol.~7, no.~1, pp. 113--125, Mar. 2006.

\bibitem{Jing2008}
S.~Jing, D.~N.~C. Tse, J.~B. Soriaga, J.~Hou, J.~E. Smee, and R.~Adovani,
  ``Multicell downlink capacity with coordinated processing,'' \emph{{EURASIP}
  Journal of Wireless Communications and Networking}, vol. 2008, p. 19 pages,
  2008.

\bibitem{Ng2005}
B.~L. Ng, J.~S. Evans, S.~V. Hanly, and D.~Aktas, ``Transmit beamforming with
  cooperative base stations,'' in \emph{Proc. of the {IEEE} International
  Symposium on Information Theory}, Sept. 2005, pp. 1431 -- 1435.

\bibitem{Ng2008}
------, ``Distributed downlink beamforming with cooperative base stations,''
  \emph{{IEEE} Transactions on Information Theory}, vol.~54, no.~12, pp.
  5491--5499, Dec. 2008.

\bibitem{Somekh2006}
O.~Somekh, O.~Simeone, Y.~Bar-Ness, and A.~M. Haimovich, ``{CTH11-2}:
  {D}istributed multi-cell zero-forcing beamforming in cellular downlink
  channels,'' in \emph{Proc. of the {IEEE} Global Telecommunications
  Conference}, Nov. 2006, pp. 1--6.

\bibitem{Ekbal2005}
J.~Ekbal, A.;~Cioffi, ``Distributed transmit beamforming in cellular networks -
  a convex optimization perspective,'' in \emph{Proc. of the {IEEE}
  International Conference on Communications}, vol.~4, May 2005, pp.
  2690--2694.

\bibitem{Lee2008}
B.~O. Lee, H.~W. Je, I.~Sohn, O.-S. Shin, and K.~B. Lee, ``Interference-aware
  decentralized precoding for multicell {MIMO TDD} systems,'' in \emph{Proc. of
  {IEEE} Global Telecommunications Conference}, Dec. 2008, pp. 1--5.

\bibitem{Lee2009}
B.~O. Lee, H.~W. Je, O.-S. Shin, and K.~B. Lee, ``A novel uplink {MIMO}
  transmission scheme in a multicell environment,'' \emph{{IEEE} Transactions
  on Wireless Communications}, vol.~8, no.~10, pp. 4981--4987, Opt. 2009.

\bibitem{Chae2009}
C.-B. Chae, I.~Hwang, {R.~W.~Heath, Jr.}, and V.~Tarokh, ``Interference
  aware-coordinated beamforming system in a two-cell environment,''
  \emph{{IEEE} Journal of Selected Areas in Communications}, 2009.

\bibitem{Somekh2007}
O.~Somekh, B.~M. Zaidel, and S.~Shamai, ``Sum rate characterization of joint
  multiple cell-site processing,'' \emph{{IEEE} Transactions on Information
  Theory}, vol.~53, no.~12, pp. 4473--4497, Dec. 2007.

\bibitem{3GPPAdv}
{3GPP, TR 36.814}, ``Further advancements for {E-UTRA}; physical layer
  aspects,'' Tech. Rep., 2008.

\bibitem{Love2008}
D.~J. Love, {R. W. Heath, Jr.}, V.~K.~N. Lau, D.~Gesbert, B.~D. Rao, and
  M.~Andrews, ``An overview of limited feedback in wireless communication
  systems,'' \emph{{IEEE} Journal on Selected Areas in Communications},
  vol.~26, no.~8, pp. 1341 -- 1365, Oct. 2008.

\bibitem{Costa1983}
M.~Costa, ``Writing on dirty paper,'' \emph{{IEEE} Transactions on Information
  Theory}, vol.~29, no.~3, pp. 439--441, May 1983.

\bibitem{Wyner1994}
A.~D. Wyner, ``Shannon-theoretic approach to a {G}aussian cellular
  multiple-access channel,'' \emph{{IEEE} Transactions on Information Theory},
  vol.~40, pp. 1713 -- 1727, Nov. 1994.

\bibitem{Mielczarek2008}
B.~Mielczarek and W.~A. Krzymien, ``Comparison of partial {CSI} encoding
  methods in multi-user {MIMO} systems,'' \emph{Wireless Personal
  Communications}, May 2008.

\bibitem{Zhang2009a}
J.~Zhang, {R. W. Heath, Jr.}, M.~Kountouris, and J.~G. Andrews, ``Mode
  switching for the {MIMO} broadcast channel based on delay and channel
  quantization,'' \emph{{EURASIP} Journal on Advances in Signal Processing,
  special issue on Multiuser Limited Feedback}, vol. 2009, 2009.

\bibitem{Trivellato2008}
M.~Trivellato, S.~Tomasin, and N.~Benvenuto, ``Channel quantization and
  feedback optimization in multiuser {MIMO-OFDM} downlink systems,'' in
  \emph{Proc. of {IEEE} Global Telecommunications Conference}, Mar 2008, pp.
  1--5.

\bibitem{Santipach2004}
W.~Santipach and M.~Honig, ``Asymptotic capacity of beamforming with limited
  feedback,'' in \emph{{IEEE} International Symposium on Information Theory},
  Jun./Jul. 2004, pp. 290--295.

\bibitem{Santipach2005}
------, ``Signature optimization for {CDMA} with limited feedback,''
  \emph{{IEEE} Transactions on Information Theory}, vol.~51, no.~10, pp. 3475
  -- 3492, 2005.

\bibitem{LF-Assumption1}
W.~Choi, A.~Forenza, J.~G. Andrews, and {R. W. Heath, Jr.}, ``Opportunistic
  space-division multiple access with beam selection,'' \emph{{IEEE}
  Transactions on Communications}, vol.~55, no.~12, pp. 2371--2380, Dec. 2007.

\bibitem{LF-Assumption2}
M.~Sharif and B.~Hassibi, ``On the capacity of {MIMO} broadcast channels with
  partial side information,'' \emph{{IEEE} Transactions on Information Theory},
  vol.~51, no.~2, pp. 506Ð--522, Feb. 2005.

\bibitem{Huang2009}
K.~Huang, J.~G. Andrews, and {R. W. Heath, Jr.}, ``Performance of orthogonal
  beamforming for {SDMA} with limited feedback,'' \emph{{IEEE} Transactions on
  Vehicular Technology}, vol.~57, no.~5, pp. 1959--1975, 2009.

\bibitem{Mondal2004}
B.~Mondal and {R. W. Heath, Jr.}, ``An upper bound on {SNR} for limited
  feedback {MIMO} beamforming systems,'' in \emph{Proc. of the {IEEE}
  Information Theory Workshop}, October 2004, pp. 408--412.

\bibitem{Borga1998}
M.~Borga, ``Learning multidimensional signal processing,'' Ph.D. dissertation,
  Link¬oping University, Sweden, SE-581 83 Link¬oping, Sweden, 1998.

\bibitem{Cadambe2008}
V.~R. Cadambe and S.~A. Jafar, ``Interference alignment and degrees of freedom
  of the {K}-user interference channel,'' \emph{{IEEE} Transactions on
  Information Theory}, vol.~54, pp. 3425--3441, Aug. 2008.

\bibitem{Sadek2007}
M.~Sadek, A.~Tarighat, and A.~H. Sayed, ``A leakage-based precoding scheme for
  downlink multi-user {MIMO} channels,'' \emph{{IEEE} Transactions on Wireless
  Communications}, vol.~6, no.~5, pp. 1711--1721, May 2007.

\bibitem{Lim2009}
M.~C.~H. Lim, D.~C. McLernon, and M.~Ghogho, ``Weighted harmonic mean {SINR}
  maximization for the {MIMO} downlink,'' in \emph{{IEEE} International
  Conference on Acoustics, Speech and Signal Processing,}, Apr. 2009, pp.
  2381--2384.

\bibitem{Gokbayrak1999}
K.~Gokbayrak and C.~G. Cassandras, ``Stochastic discrete optimization using a
  surrogate problem methodology,'' in \emph{Proc. of the {IEEE} Conference on
  Decision \& Control}, Dec. 1999, pp. 1779--1784.

\bibitem{Murota2003}
K.~Murota, \emph{Discrete Convex Analysis}.\hskip 1em plus 0.5em minus
  0.4em\relax Society for Industrial and Applied Mathematics, 2003.

\bibitem{Au-Yeung2007}
C.~K. Au-Yeung and D.~J. Love, ``On the performance of random vector
  quantization limited feedback beamforming in a {MISO} system,'' \emph{{IEEE}
  Transactions on Wireless Communications}, vol.~6, no.~2, pp. 458--462, Feb.
  2007.

\bibitem{intSeriesBook}
I.~S. Gradshteyn and I.~M. Ryzhik, \emph{Tables of Integrals, Series and
  Products}, 7th~ed.\hskip 1em plus 0.5em minus 0.4em\relax Academic Press,
  2007.

\bibitem{IntTheoryBook}
M.~Vath, \emph{Integration Theory: A Second Course}.\hskip 1em plus 0.5em minus
  0.4em\relax World Scientific Publishing Company, 2002.

\end{thebibliography}
\newpage

\centering{\large{\textbf{Figures}}

\begin{figure} [h!]
  \centering
  \includegraphics[width=6.25in]{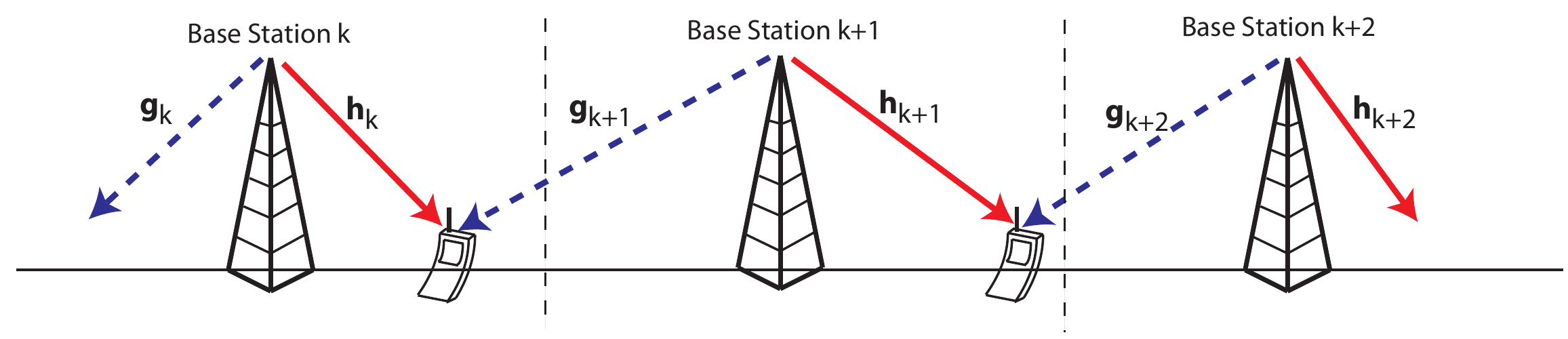}
  \caption{Pictorial depiction of the Wyner model (described in Section~\ref{sec:SysModel}). The solid line represents the desired signal, while the dashed line represents the interfering signal.}\label{fig:WynerModel}
\end{figure}




\begin{figure} [h!]
  \centering
  \includegraphics[width=6in]{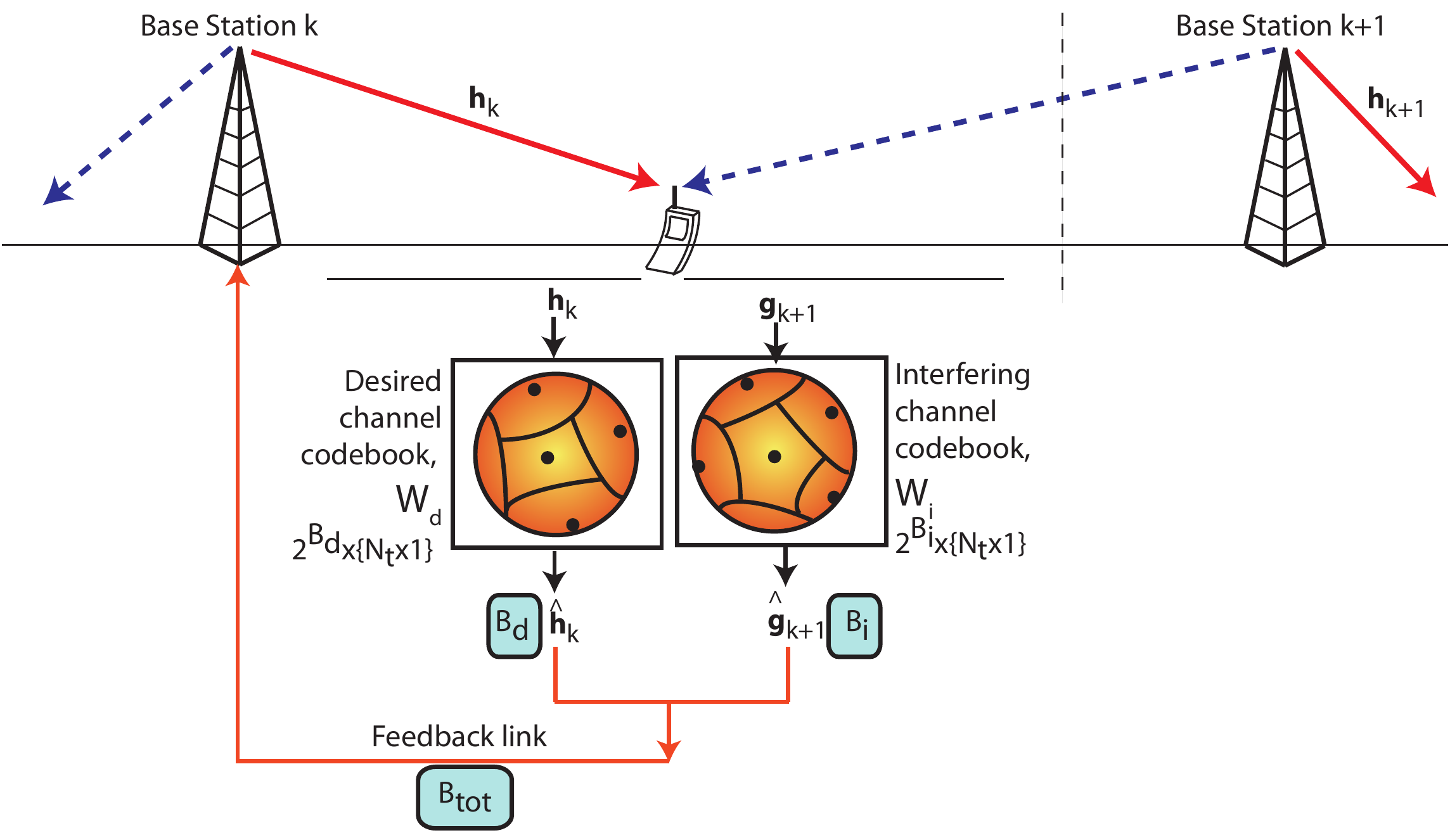}
  \caption{The limited feedback model, described in Section~\ref{sec:BFVec_PartialCSI}, to feedback quantized CSI of the desired and interfering channels using two separate codebooks.}\label{fig:LFModel}
\end{figure}

\begin{figure} [h!]
  \centering
  \includegraphics[width=4.5in]{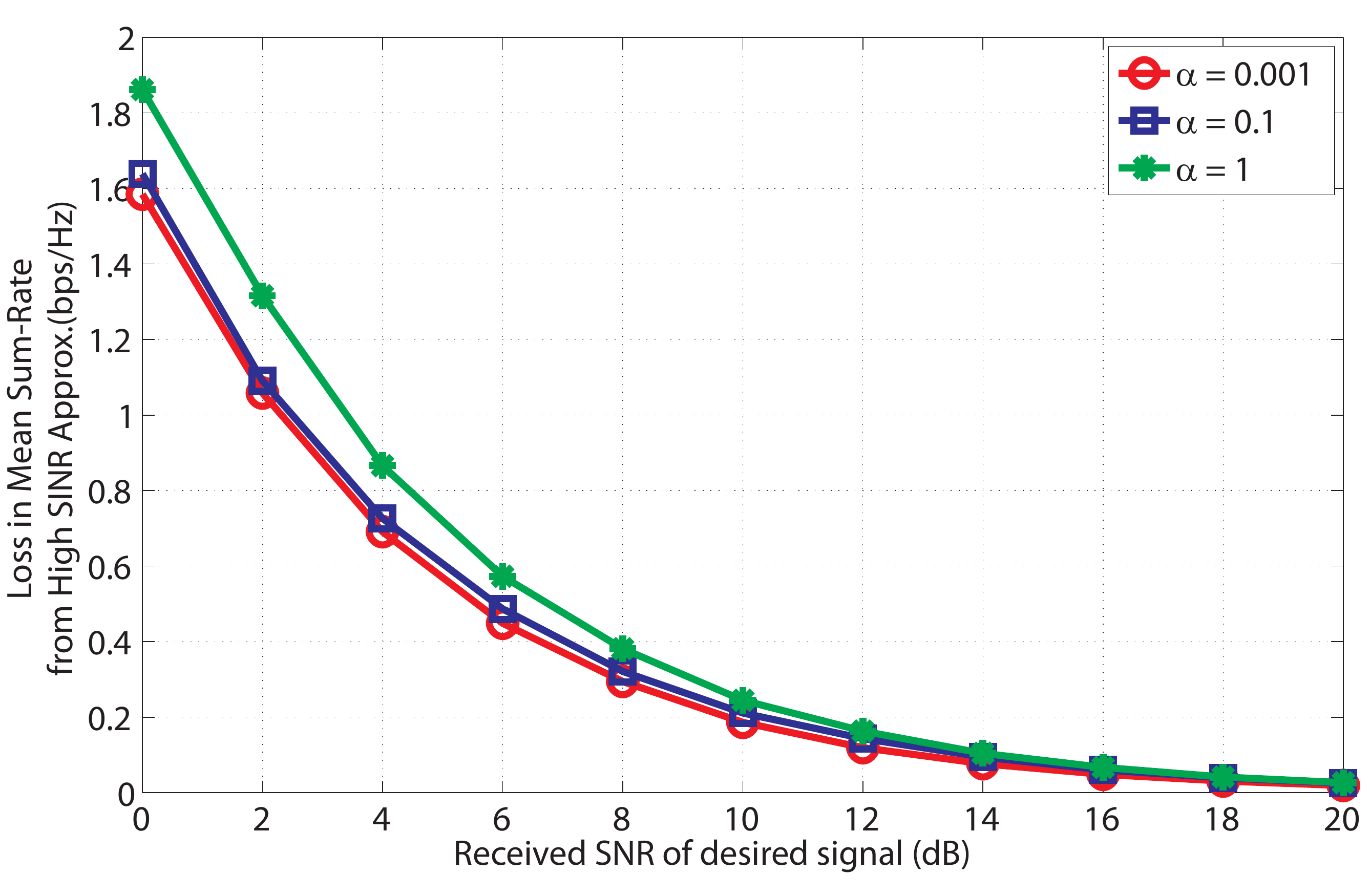}
  \caption{Comparison of the actual and high SINR approximated sum-rates (in \eqref{eqn:SumRate} and \eqref{eqn:SumRate2}, respectively) for $\alpha = \{0.001,0.1,1\} $, when $\Nt = 4$, $\rho_{\rm(d)} = [0,20]~dB$ and $K=4$ for full CSI.}\label{fig:FullCSI_HighSINRApprox}
\end{figure}

\begin{figure} [h!]
  \centering
  \includegraphics[width=4.5in]{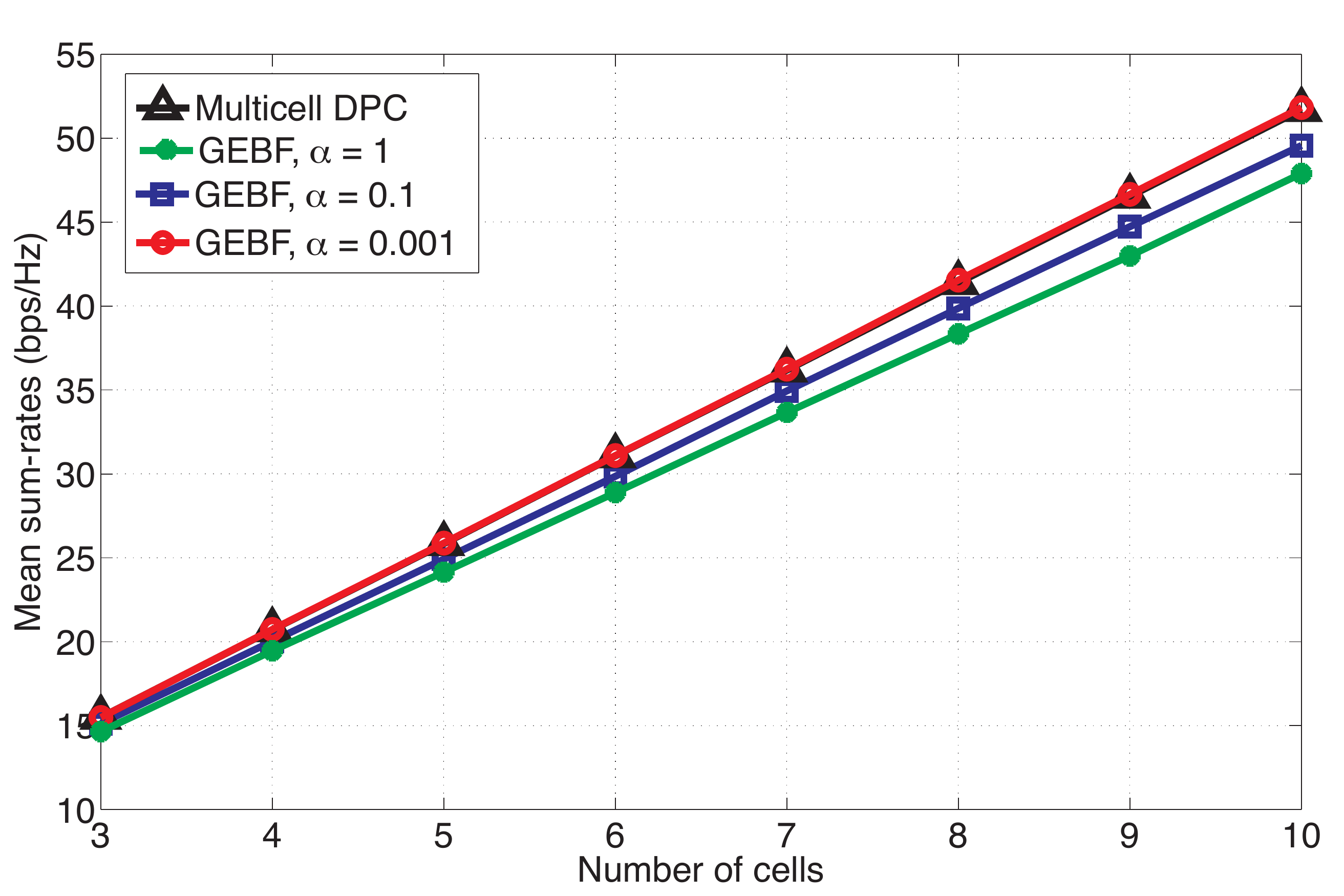}
  \caption{Sum-rates as a function of the number of cells ($K$) for $\alpha = \{0.001,0.1,1\} $, when $\Nt = 4$, $\rho_{\rm(d)} = 10~dB$, for GEBF with full CSI.}\label{fig:SumRate_GEBF_DPC_IncrK}
\end{figure}



\begin{figure} [h!]
  \centering
  \includegraphics[width=4.5in]{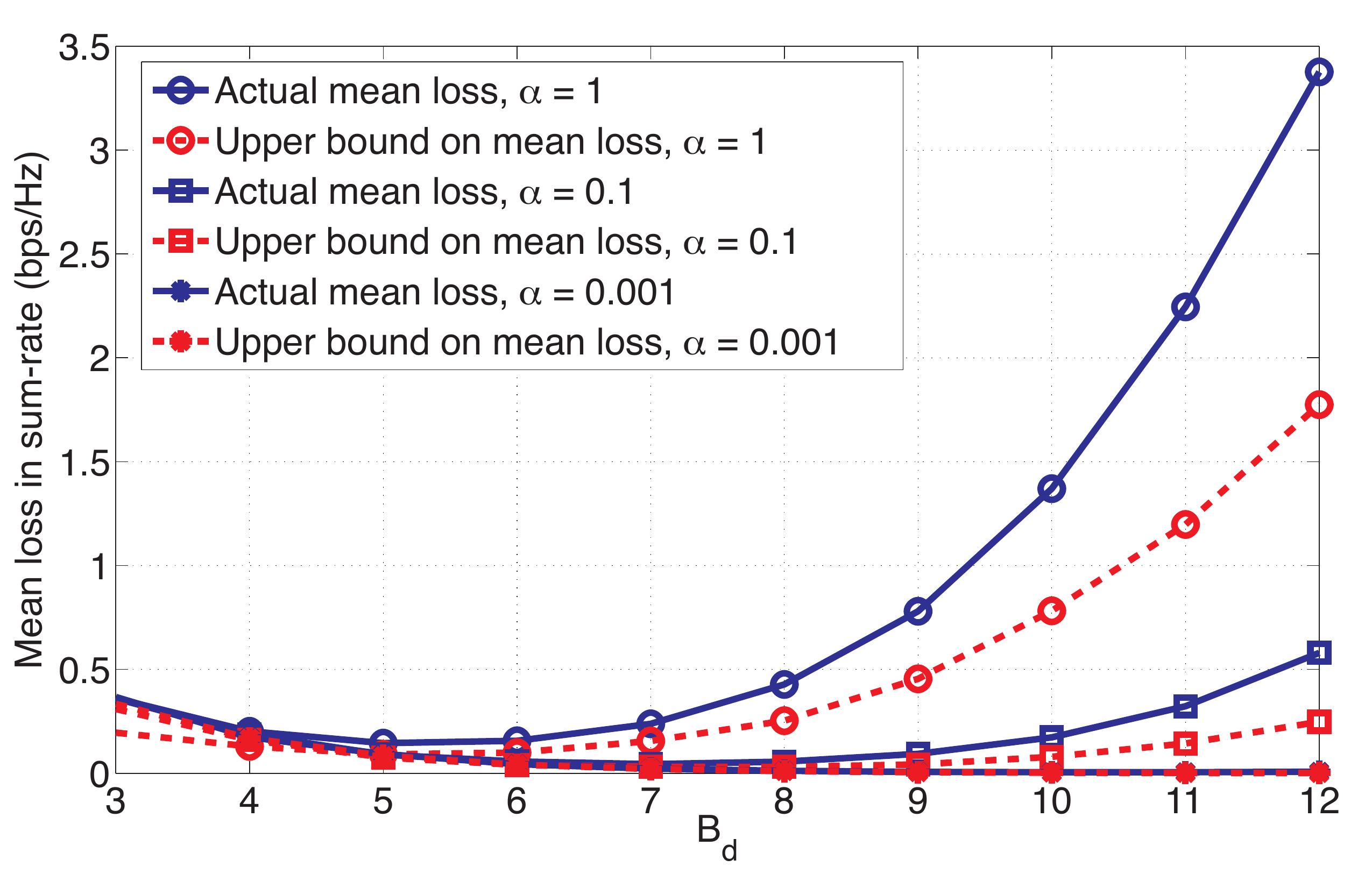}
  \caption{Mean loss in sum-rate for $\alpha = \{10.001,0.1,1\}$, $\Nt = K = 2$, $\rho_{\rm(d)} = 10~dB$ and $\Bt = 15$. }\label{fig:UppBndComp}
\end{figure}


\begin{figure} [h!]
  \centering
  \includegraphics[width=4.5in]{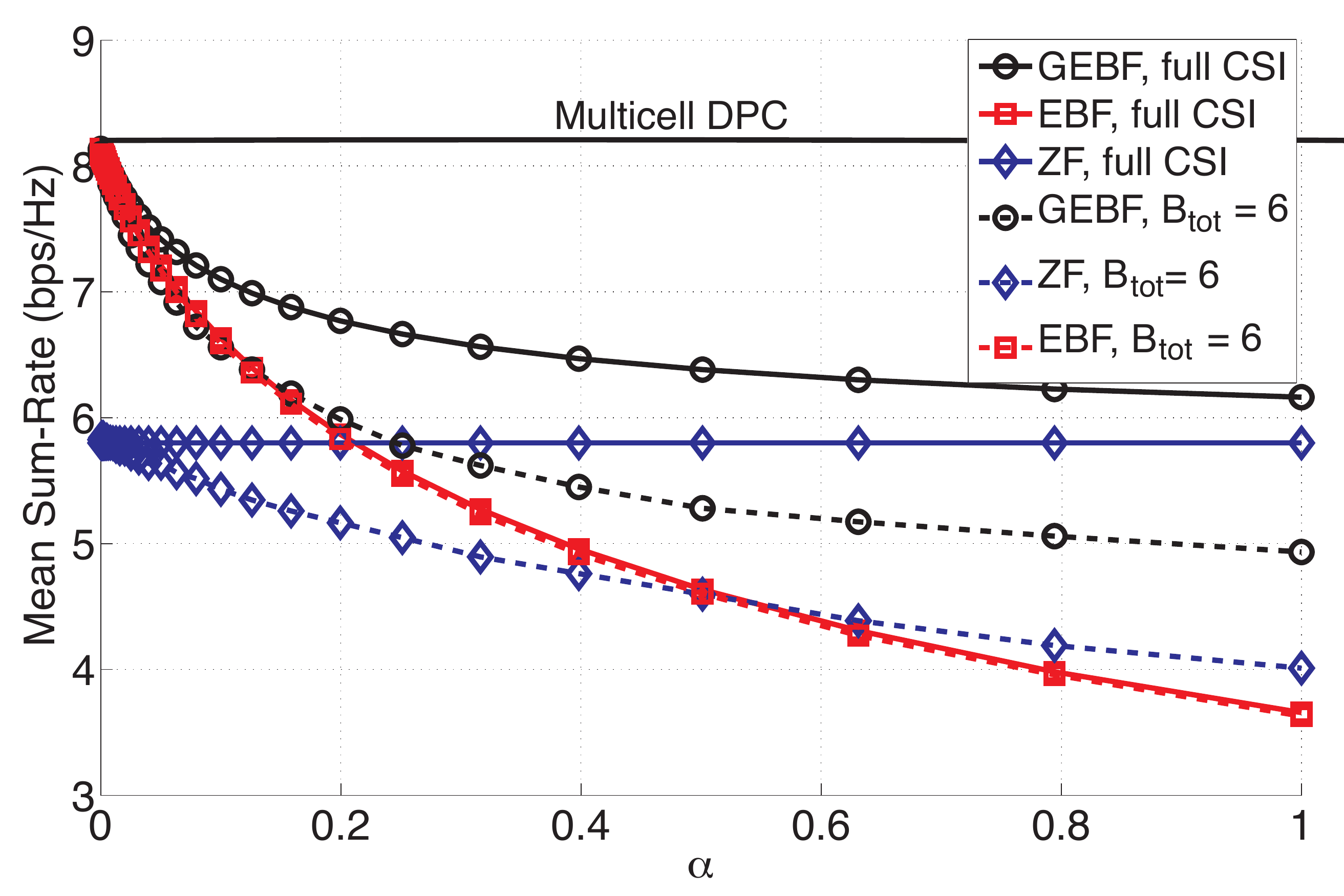}
  \caption{Comparison of the mean sum-rates obtained using the proposed beamforming strategy, non-cooperative eigenbeamforming and (cooperative) zero forcing, for $\rho_{\rm(d)} = 10~dB$, $\Nt = K = 2$ and $\Bt = 6$. }\label{fig:CompTxStrat}
\end{figure}

\begin{figure} [h!]
  \centering
  \includegraphics[width=4.5in]{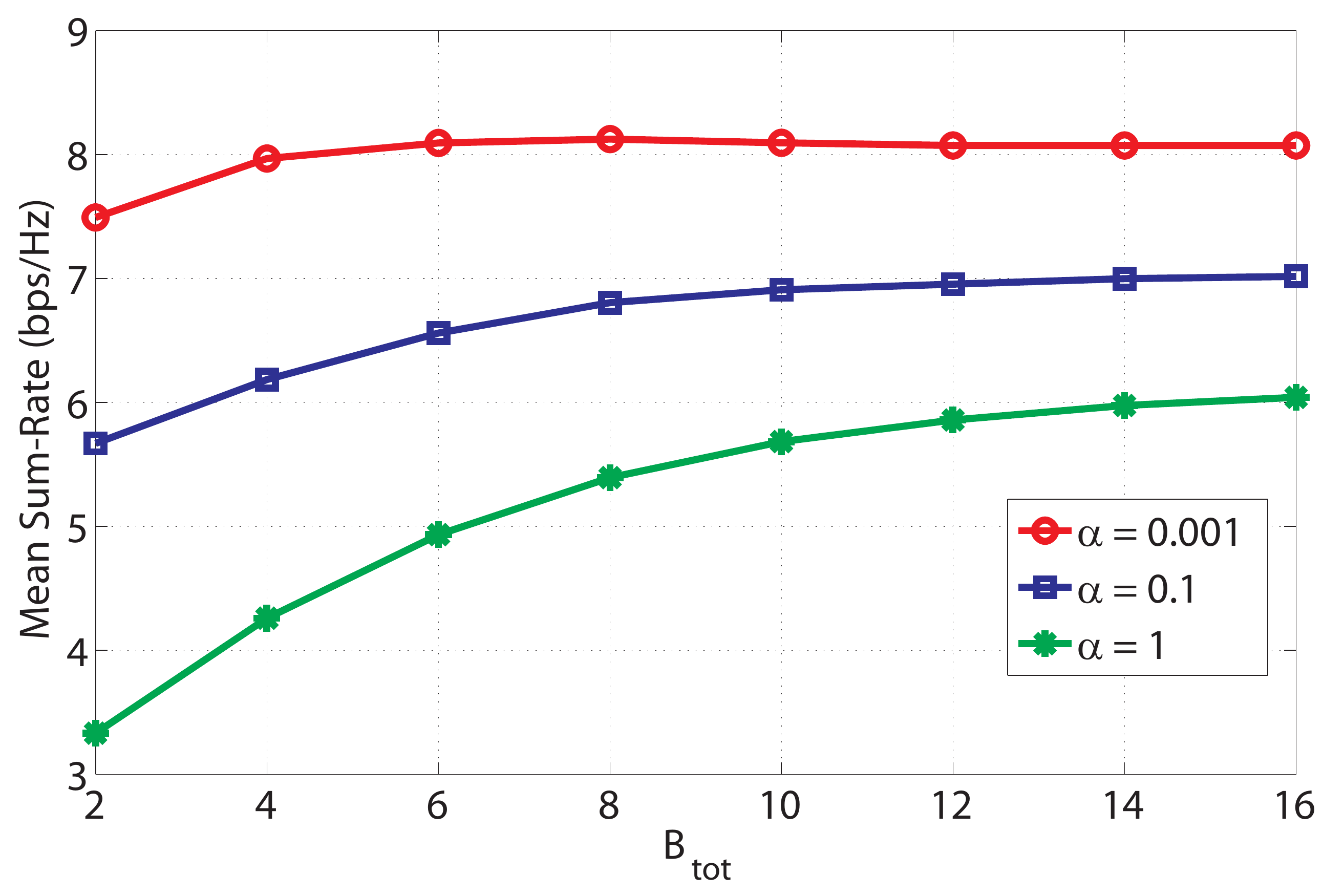}
  \caption{Comparison of the mean sum-rates obtained using GEBF for $\alpha = \{0.001,0.1,1\}$ ad a function of $\Bt$, for $\rho_{\rm(d)} = 10~dB$, $\Nt = K = 2$. }\label{fig:SumRate_Vs_Btot_3Strat}
\end{figure}

\begin{figure} [h!]
  \centering
  \includegraphics[width=4.5in]{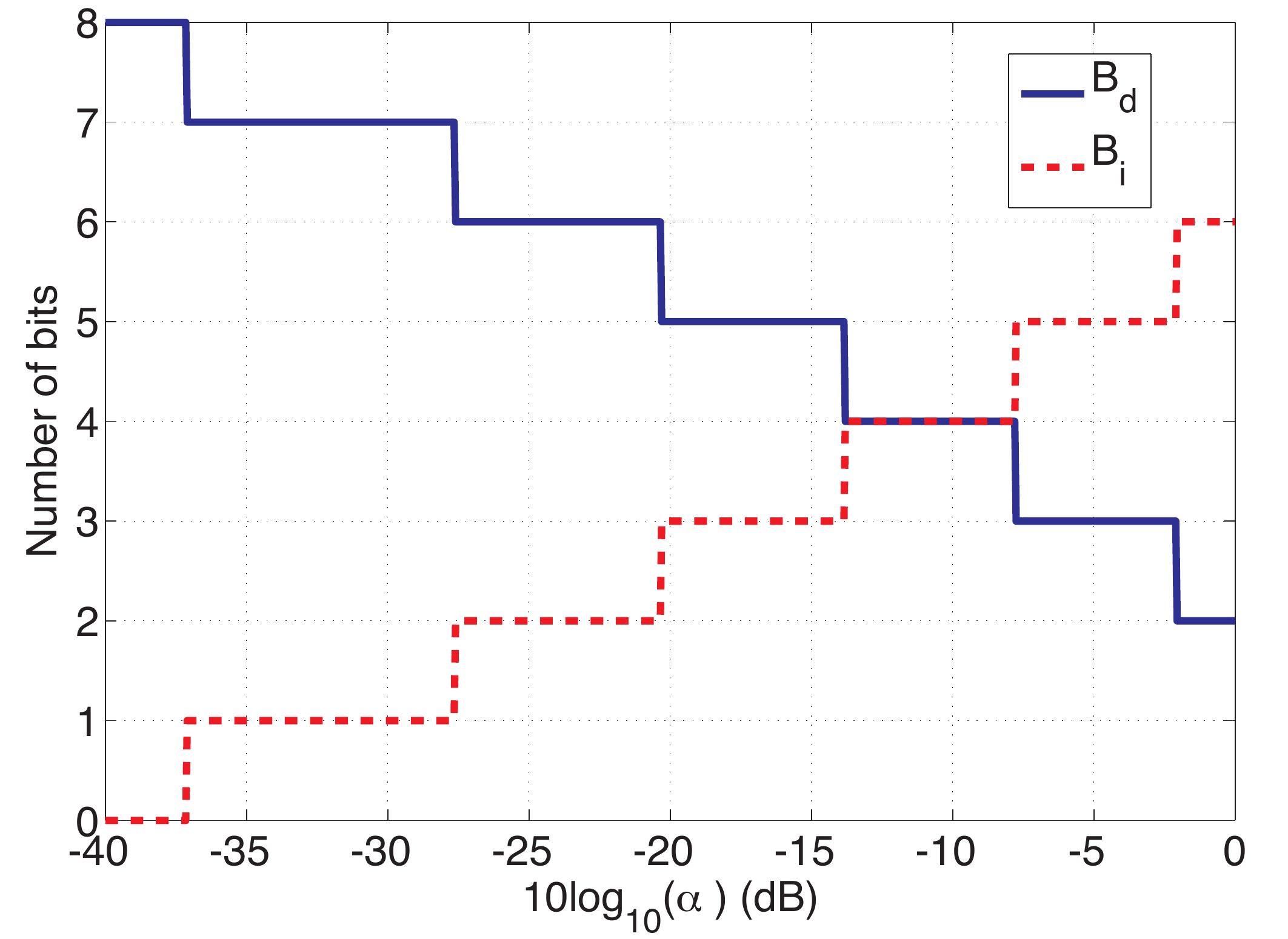}
  \caption{$(B_d,B_i)$ partitioning for, for $\rho_{\rm(d)} = 10~dB$, $\Nt = K = 2$ and $\Bt=8$. }\label{fig:BdBi_Vs_LogAlpha}
\end{figure}

\begin{figure} [h!]
  \centering
  \includegraphics[width=5in]{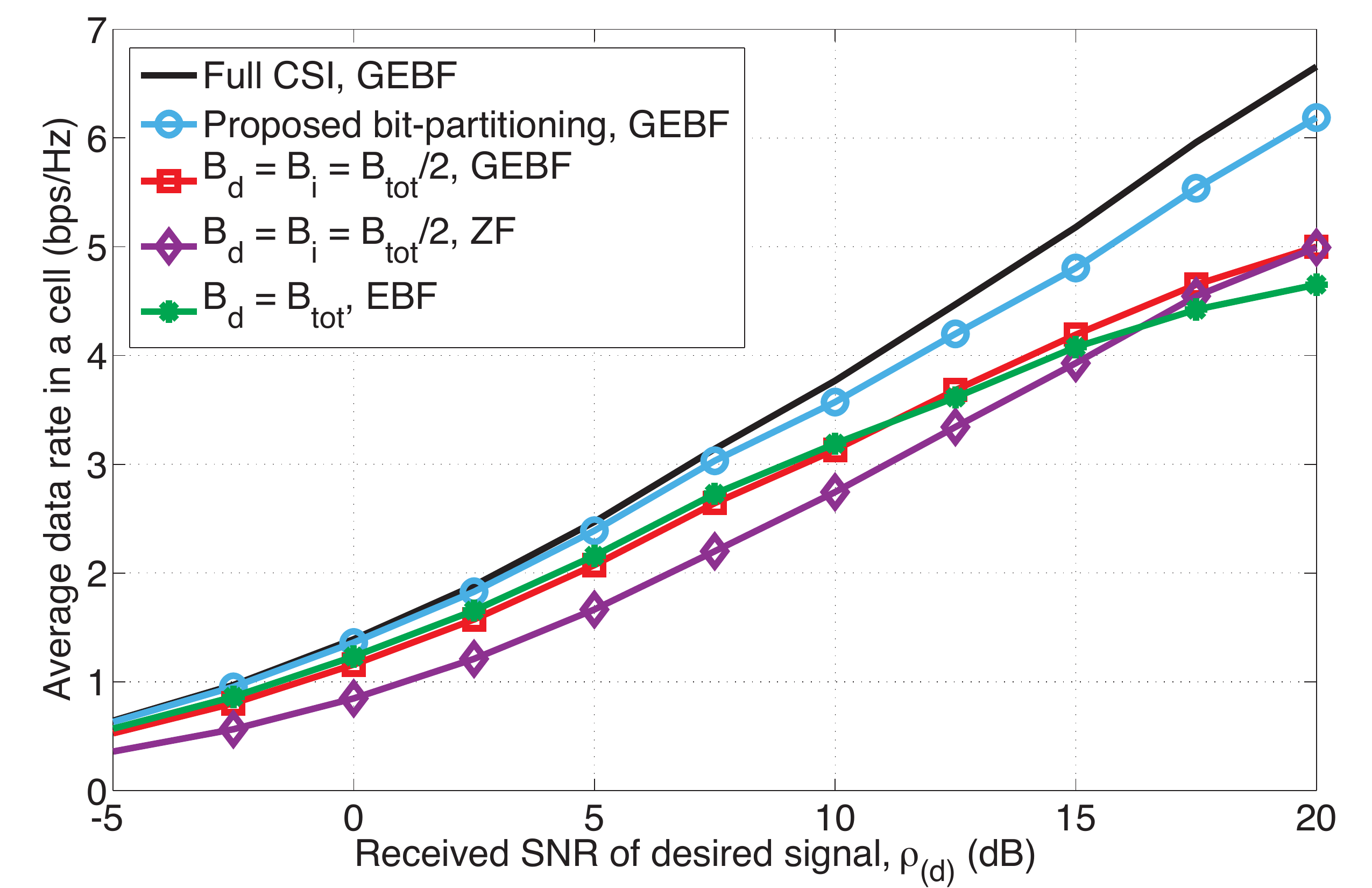}
  \caption{Average data rate in a cell as a function of the received desired signal power for $\Nt = K = 2$ and $\Bt=6$. }\label{fig:Bits6_diffAlphas}
\end{figure}

\end{document}